\documentclass[aps,twocolumn,pra,superscriptaddress,letterpaper,nofootinbib,longbibliography]{revtex4-1}
\usepackage{amssymb}
\usepackage{physics}
\usepackage{mathtools}
\usepackage{upgreek}
\usepackage{booktabs}
\AtBeginDocument{
\heavyrulewidth=.08em
\lightrulewidth=.05em
\cmidrulewidth=.03em
\belowrulesep=.65ex
\belowbottomsep=0pt
\aboverulesep=.4ex
\abovetopsep=0pt
\cmidrulesep=\doublerulesep
\cmidrulekern=.5em
\defaultaddspace=.5em
}
\usepackage{txfonts}
\usepackage[pdftex]{graphicx}
\usepackage[usenames, dvipsnames, svgnames, table]{xcolor}
\usepackage[colorlinks=true, citecolor=Blue,linkcolor=BrickRed, urlcolor=magenta]{hyperref}
\usepackage{bm}
\usepackage{placeins} 
\usepackage{setspace}
\usepackage[final]{changes}

\newcommand{\DC}{\text{DC}}
\newcommand{\beat}{b}
\newcommand{\sn}{s}
\newcommand{\RMS}[1]{\text{RMS}\left\{#1\right\}}
\newcommand{\sumqpd}{\sum\limits_{k=A}^{D}}

\begin{document}

\title{Tracking length and differential wavefront sensing signals from quadrant photodiodes in heterodyne interferometers with digital phase-locked loop readout}

\author{Gerhard Heinzel}
\email{gerhard.heinzel@aei.mpg.de}
\affiliation{Albert-Einstein Institut (Max-Planck Institut f\"ur Gravitationsphysik),\\
Callinstrasse 38, D-30167 Hannover, Germany}

\author{Miguel Dovale \'Alvarez}
\email{miguel.dovale@aei.mpg.de}
\affiliation{Albert-Einstein Institut (Max-Planck Institut f\"ur Gravitationsphysik),\\
Callinstrasse 38, D-30167 Hannover, Germany}

\author{Alvise Pizzella}
\affiliation{Albert-Einstein Institut (Max-Planck Institut f\"ur Gravitationsphysik),\\
Callinstrasse 38, D-30167 Hannover, Germany}

\author{Nils Brause}
\affiliation{Albert-Einstein Institut (Max-Planck Institut f\"ur Gravitationsphysik),\\
Callinstrasse 38, D-30167 Hannover, Germany}

\author{Juan Jos\'e Esteban Delgado}
\affiliation{Albert-Einstein Institut (Max-Planck Institut f\"ur Gravitationsphysik),\\
Callinstrasse 38, D-30167 Hannover, Germany}

\begin{abstract}
We propose a method to track signals from quadrant photodiodes (QPD) in heterodyne laser interferometers that employ digital phase-locked loops for phase readout. Instead of separately tracking the four segments from the QPD and then combining the results into length and Differential Wavefront Sensing (DWS) signals, this method employs a set of coupled tracking loops that operate directly on the combined length and angular signals. Benefits are increased signal-to-noise ratio in the loops and the possibility to adapt the loop bandwidths to the different dynamical behavior of the signals being tracked, which now correspond to physically meaningful observables. We demonstrate an improvement of up to 6\,dB over single-segment tracking, which makes this scheme an attractive solution for applications in precision inter-satellite laser interferometry in ultra-low-light conditions.
\end{abstract}

\maketitle

\noindent Phys. Rev. Applied \textbf{14}, 054013 $-$ 6 November 2020 \href{https://doi.org/10.1103/PhysRevApplied.14.054013}{DOI:10.1103/PhysRevApplied.14.054013}

\vspace{1cm}

\section{Introduction}

Laser interferometry is a powerful method to measure tiny distance variations as changes of optical pathlengths. When the optical pathlengths cannot be kept constant to within a small fraction of a wavelength, heterodyne interferometry is frequently applied, i.e., by interfering two laser beams that have a finite frequency difference. The interference pattern is recorded with a photodiode, which produces a photocurrent with a sinusoidal component at the heterodyne frequency. Differential changes of optical pathlength are then converted into phase changes of that sinusoidal beatnote, which are measured with a phasemeter instrument. Heterodyne laser interferometers have been employed with great success in space-based gravity missions~\cite{LTP2, PhysRevLett.123.031101}, which require high precision displacement sensors with a large dynamic range.

Several techniques exist to perform this phase measurement with sub-picometer precision in the millihertz frequency band as required for applications in gravimetry, gradiometry, and gravitational-wave detection~\cite{PMLTP2, PMLTP1, PMCROSS, PMZERO, PMLISA1, PMLISA2, PMLISA3, PMLISA4, PMLISA5, PMLRI}. When the heterodyne frequency changes with time, as e.g., over inter-spacecraft optical links such as the LISA mission~\cite{LISA, L3Proposal} or the Laser Ranging Interferometer (LRI) on GRACE Follow-On~\cite{LRI, LRIRES}, the usual technique is to employ a Digital Phase-Locked Loop (DPLL)~\cite{PMLISA1, PMLISA2, PMLISA3, PMLISA4, PMLISA5, PMLRI, OLI} architecture to track the amplitude, frequency and phase of the beatnote, even when the received beam is sensed at sub-picowatt power level due to the beam divergence over the large propagation distances typical of inter-satellite laser links.

\begin{figure}[t]
\begin{centering}
\includegraphics{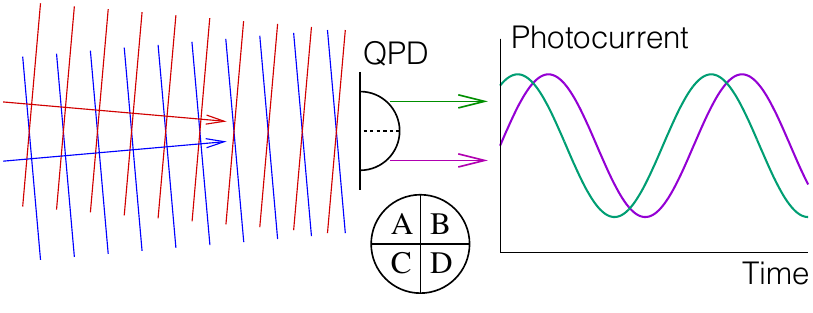}
\caption{\label{fig:dws} Illustration of the Differential Wavefront Sensing method. Two waves impinge on the surface of a quadrant photodiode (QPD) at an angle. The upper and lower segments of the QPD thus measure different phase offsets between the interfering wavefronts. In the case of heterodyne interference, this angular misalignment between wavefronts manifests as shifts of the phase of the measured photocurrent beatnote signals.}
\end{centering}
\end{figure}

In many applications of laser interferometers, such as in gravitational reference sensors~\cite{LTP1}, spacecraft attitude control~\cite{PhysRevD.99.082001}, or precision laser beam pointing to remote spacecraft~\cite{Koch2018}, it is beneficial or even essential to determine not only the relative optical pathlength change between the two interfering beams, but also the angle between their wavefronts, since the latter is a very sensitive measurement of misalignments in the optical system. The standard technique to achieve this angular measurement is Differential Wavefront Sensing (DWS)~\cite{DWS1,DWS2}, which uses a quadrant photodiode (QPD) to detect the interference pattern.  The average phase over the four segments represents the length signal, while the difference between left and right or top and bottom represent horizontal and vertical misalignments, respectively (see Figure~\ref{fig:dws}).

In previous implementations, the phase measurement is applied separately to each segment of the diode, and the results are then combined. This paper describes a method to track the phases of the beatnotes from the four segments of a quadrant photodiode with DPLLs. In Section~\ref{sec:dpll}, the function of a DPLL as a phasemeter core is summarized. The standard application of four independent DPLLs to the segments of a QPD is described in Section~\ref{sec:dwsold}, and the proposed scheme in Section~\ref{sec:new}. A noise analysis of the new technique, with a comparison against the conventional method, is reported in Section~\ref{sec:tests}, followed by a conclusion in Section~\ref{sec:conclusion}.

\section{Phase measurement with Digital Phase-Locked loops}
\label{sec:dpll}
\begin{figure}[t]
\begin{centering}
\includegraphics{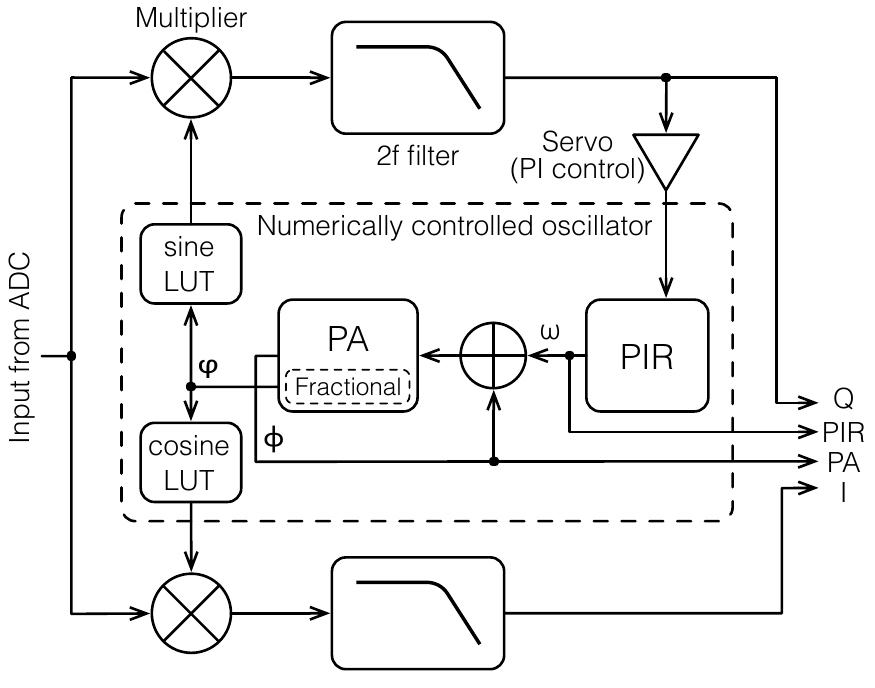}
\caption{ \label{fig:dpll} Functional blocks of a Digital Phase-Locked Loop (DPLL). The input signal from the ADC is separately mixed with a sine and a cosine delivered by the numerically controlled oscillator (NCO), yielding the `in Quadrature' $Q$ and an `in Phase' $I$ signals. The second harmonic of the beatnote frequency $2f$ is suppressed by low-pass filtering these signals. The $Q$ signal is then used as error signal for the servo. The output of the PI controller gives the instantaneous signal frequency, and is stored in the Phase Increment Register (PIR). This is then integrated and stored in the Phase Accumulator (PA), which is fed to the sine and cosine look up tables (LUT) that close the loop.}
\end{centering}
\end{figure}

The principle of a DPLL is to generate a digital sine wave in a Numerically Controlled Oscillator (NCO), and to make it track the incoming sinusoidal beatnote signal in frequency and phase. After appropriate signal conditioning, the incoming signal is first digitized in an Analog-to-Digital converter (ADC), and all remaining processing happens in the digital domain, typically implemented in a Field-Programmable Gate Array (FPGA) for the tracking part, see Figure~\ref{fig:dpll}.

The NCO consists of a Phase Increment Register (PIR) that represents the instantaneous signal frequency, a Phase Accumulator (PA) which holds the integral of the frequency, i.e.\ the instantaneous phase, and a Look-Up-Table (LUT) that converts the phase into a sine wave and optionally also a cosine wave. The ADC and all digital blocks are driven synchronously from a common clock, which sets the reference for any single phase measurement.

In order to make the NCO sine signal track the incoming signal,
both are mixed in a multiplier that acts as phase detector, and the phase deviation thus measured is used as error signal in a servo loop. When the loop is closed and locked, both the incoming and the NCO sine signal have the same frequency, and their phase is shifted by 90$^\circ$, such that their product, the error signal, has zero average. Incoming and NCO sine signal are said to be ``in Quadrature'' (denoted by `Q'). An optional second branch multiplies the incoming signal with a digital cosine signal, which is then ``In Phase'' (`I') and which can be used to measure the amplitude of the incoming signal. Lowpass filters after the mixers suppress the second harmonic of the signal frequency (`$2f$'), a by-product of the multiplication, in order to prevent it from circulating around the loop in an undesired non-linear process. The primary achievement of such a DPLL is that frequency and phase now exist in digital form in the PIR and PA registers, respectively, within the NCO from where they can be directly read out.

More specifically, the PIR holds the instantaneous signal frequency $\omega$ in units of cycles per clock period, with ${0<\omega<0.5}$. It slowly varies as the input signal frequency changes. It is integrated in the PA which always has a fractional part $\varphi$, with ${0\le\varphi <1\,{\rm cycles}}$, which is used by the LUTs. It follows a rapid sawtooth function. In most cases, the integer number of cycles (wavelengths) must also be tracked. This can be achieved by including extra bits in the PA that represent the integer number of cycles. We denote that extended PA by $\Phi$, with 
\begin{equation}
\varphi=\Phi\;{\rm mod}\; 1
\end{equation}
simply being the fractional part of $\Phi$. The total phase $\Phi$ is an ever increasing ramp. Instead of using extra bits in the PA, the total phase can in principle also be reconstructed by integrating $\omega$ externally.

Among the many performance parameters of a DPLL, most important here is the ability to continuously track the input signal without cycle slips, i.e.\ integer errors in $\Phi$ caused by excess noise in the input signal. That robustness can be optimized by carefully adapting the servo gain to the signal dynamics and enhancing the signal-to-noise ratio (SNR) of the input signal~\cite{PMLISA5,OLI,GARDNER}.

\begin{figure}
\begin{centering}
\includegraphics{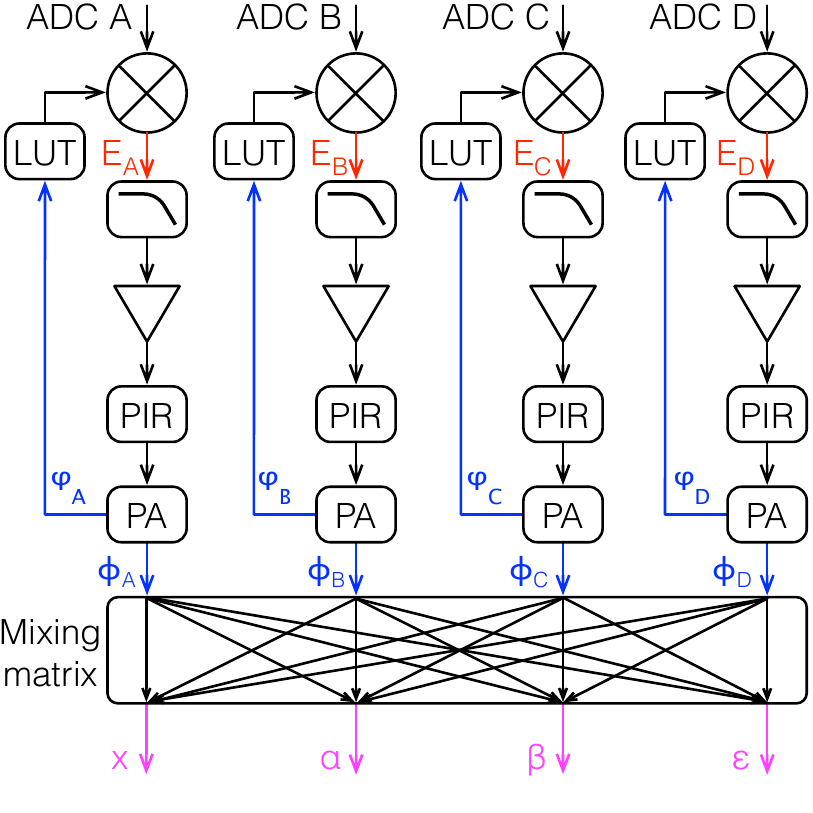}
\caption{ \label{fig:dwsold} Conventional readout scheme for a quadrant photodiode. Each DPLL processes 1/4 of the total signal power which nevertheless has the full dynamics and noise of the length signal.}
\end{centering}
\end{figure}

\section{Differential Wavefront Sensing with DPLL phasemeters}
\label{sec:dwsold}

If a QPD is used in order to implement DWS, the standard procedure is to connect four independent DPLLs to the four segments $A$, $B$, $C$ and $D$ of the QPD. We call their error signals $E_A$\dots $E_D$, their PA register contents $\Phi_A \dots \Phi_D$, and the fractional parts of the latter $\varphi_A \dots \varphi_D$. They are combined to form the output signals:
\begin{align}
x &= \frac{\Phi_A+\Phi_B+\Phi_C+\Phi_D}{4},\label{eq:x}\\
\alpha &= \frac{\varphi_A-\varphi_B+\varphi_C-\varphi_D}{2},\label{eq:alpha}\\
\beta &= \frac{\varphi_A+\varphi_B-\varphi_C-\varphi_D}{2}\label{eq:beta},
\end{align}
where $x$ represents the length signal in units of the laser wavelength, and $\alpha$, $\beta$, reduced to the range ${-0.5\,{\rm cycles}\le\alpha,\beta\le+0.5\,{\rm cycles}}$, represent the horizontal and vertical tilt angles between the wavefronts, scaled by a huge factor that depends on the beam geometry~\cite{DWSFAC}. One more independent linear combination of the segment phases can be formed, which we call the {\em ellipticity}. It is rarely used but we include it here in anticipation of the next steps:
\begin{equation}
\varepsilon = \frac{\varphi_A-\varphi_B-\varphi_C+\varphi_D}{2}\label{eq:eps},
\end{equation}
The realization of this scheme is illustrated in Figure~\ref{fig:dwsold}. Combining the signals according to Equations~(\ref{eq:x})\dots(\ref{eq:beta}) is not the only possibility~\cite{SIGNALS}. In the LISA Pathfinder mission~\cite{PMLTP1,LTP1,LTP2}, for example, $x$ was computed as the argument of a complex vector formed by adding the complex amplitudes from the four segments. This differs from Equation~(\ref{eq:x}) in that the segment contributions are weighted with their respective beatnote amplitude. It is still a subject of investigation which of these methods is preferable with respect to, e.g., tilt-to-length cross-coupling. If the segment amplitudes are, however, also measured with the `I' branch of the DPLL and recorded, the results can be converted into each other in post-processing.

This standard scheme has been successfully used in many applications, notably for the continuous active beam pointing over 200\,km separation of the inter-satellite interferometer in the LRI on GRACE Follow-On~\cite{LRIRES}. There, and in other applications such as LISA, it is, however, not an optimal method:
\begin{itemize}
\item The beam divergence over the long distance together with the finite aperture of the receiving telescope results in a received beam with extremely low power (in the order of 100 pW to a few nW), and consequently in poor Signal-to-Noise ratio in the heterodyne beatnote to be tracked, which limits the achievable robustness of the tracking loops, each of which uses only 1/4 of the total signal power.
\item The length signal $x$ has a much higher dynamic range than the angular signals. The former contains the Doppler shift due to the relative spacecraft motion (typically moving at speeds of around some meters per second, corresponding to some MHz Doppler shifts for a wavelength of about 1\,\textmu m), as well as common-mode noise sources such as laser frequency noise, which largely cancel in the other three angular signals $\alpha$, $\beta$ and $\varepsilon$.
\item Nevertheless each of the four standard tracking loops contains the full length signal which sets stringent requirements on their loop bandwidths. The resulting open-loop gains are in general not optimal for the much quieter angular signals. The same holds for the $2f$ filters. 
\item The integer number of cycles, which physically exists only once for each pair of interfering beams, is represented four times in $\Phi_A \dots \Phi_D$. They should represent the same number of integer cycles and differ only by the small quantities $\alpha$, $\beta$ and $\varepsilon$. If, however,  a cycle slip occurs in only some of the four segments, the length signal $x$ computed by Equation (\ref{eq:x}) is easily messed up, and therefore the measurement degraded.
\end{itemize}

\section{Alternative architecture for tracking length and DWS signals with DPLL phasemeters}
\label{sec:new}

\begin{figure}
\begin{centering}
\includegraphics{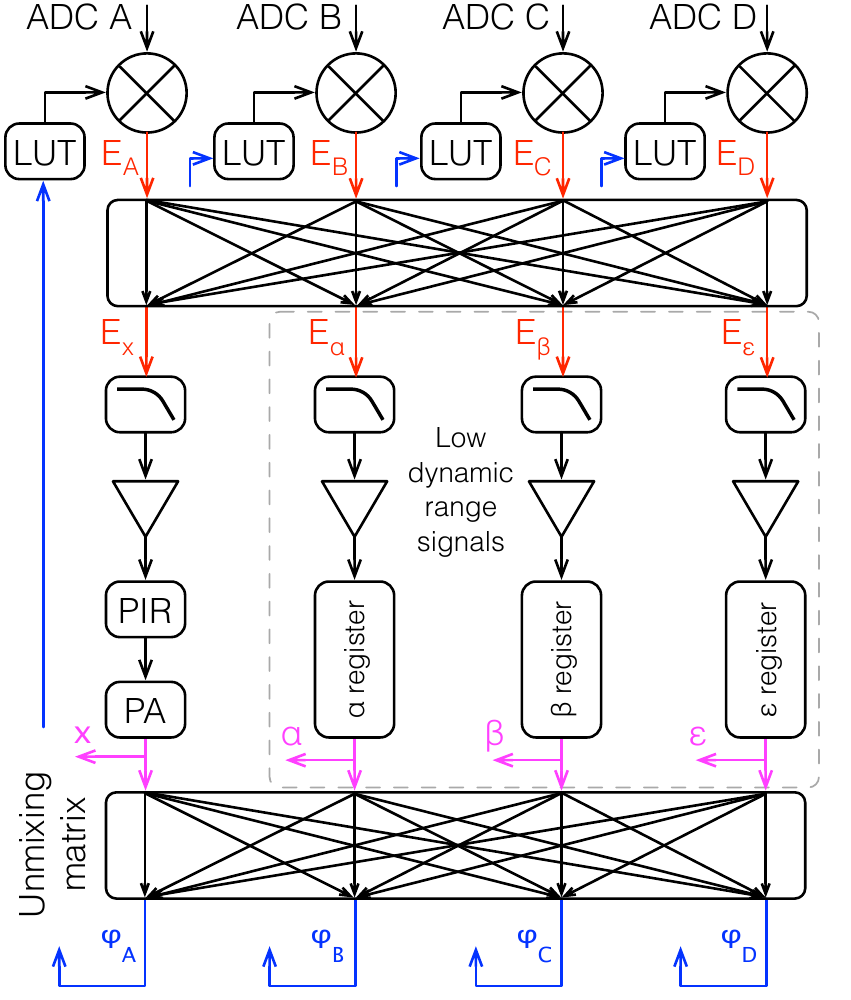}
\caption{ \label{fig:dwsnew} Proposed DPLL readout scheme for a quadrant photodiode. The error signals from the four QPD segments ${E_A \dots E_D}$ are used to compute DWS error signals ${E_x \dots E_\varepsilon}$. These are then fed to four dedicated DPLLs. The four outputs ${x \dots \varepsilon}$, now corresponding to physically meaningful observables, are then reverted back into the phase signals ${\varphi_A \dots \varphi_D}$ and used for closing the loop. Only one DPLL deals with the length signal, while the the other three loops handle signals of much smaller dynamic range (encircled). Each servo can hence be individually tuned to the particular dynamical behavior of the signal being tracked.}
\end{centering}
\end{figure}

In order to overcome the above limitations, we propose here an alternative loop topology, where the four servo loops do not act on the four segments, but on $x$, $\alpha$, $\beta$ and $\varepsilon$ instead. Error signals for these loops can be obtained from Equations similar to (\ref{eq:x}) to (\ref{eq:eps}):
\begin{align}
E_x &= \frac{E_A+E_B+E_C+E_D}{4},\label{eq:ea}\\
E_\alpha &= \frac{E_A-E_B+E_C-E_D}{2},\label{eq:eb}\\
E_\beta &= \frac{E_A+E_B-E_C-E_D}{2},\label{eq:ec}\\
E_\varepsilon &= \frac{E_A-E_B-E_C+E_D}{2}.\label{eq:ed}
\end{align}
The final registers of the four loops track $x$, $\alpha$, $\beta$ and $\varepsilon$, which directly represent the desired final output of the phasemeter. The segment phases, which are still needed for the multiplicative phase detectors, can be easily obtained by inverting Equations (\ref{eq:x}) to (\ref{eq:eps}):
\begin{align}
\varphi_A &= x+\frac{\alpha+\beta+\varepsilon}{2},\label{eq:a}\\
\varphi_B &= x+\frac{-\alpha+\beta-\varepsilon}{2},\label{eq:b}\\
\varphi_C &= x+\frac{\alpha-\beta-\varepsilon}{2},\label{eq:c}\\
\varphi_D &= x+\frac{-\alpha-\beta+\varepsilon}{2},\label{eq:d}
\end{align}
where only the fractional part of $x$ needs to be used and the results are reduced to the range $0\dots 1$ cycles. The realization of this scheme is illustrated in Figure~\ref{fig:dwsnew}.

Only $E_x$ carries the burden of the large dynamic range of the length signal, e.g., due to the Doppler effect in the orbital motion. It uses the combined signal from all four segments, each of which has been coherently demodulated with individually optimized phases. For noise sources that are uncorrelated among QPD segments, such as shot noise or electronic noise, this yields a 6\,dB improvement in the SNR, which makes the loop more robust against cycle slips. Now there is only one full NCO in the system, which produces the phase ramp $\Phi$ and its sawtooth-like fractional part $\varphi$, and in particular only one register $\Phi$ that keeps track of the integer number of cycles, which better maps the physical reality than having four registers ${\Phi_A \dots \Phi_D}$.

The signals in $E_\alpha$, $E_\beta$ and $E_\varepsilon$, on the other hand, have a much smaller dynamic range, since both the length signal as well as many correlated noise contributions, such as laser frequency noise or residual intensity noise, largely cancel out. Moreover, angular degrees of freedom typically change much slower than, e.g., spacecraft separation in applications such as the LISA mission or the GRACE Follow-On mission. The outputs of their servo loops are slowly varying numbers with a range that can be limited to e.g.\ ${-0.5\,{\rm cycles}\le\alpha,\beta,\varepsilon\le+0.5\,{\rm cycles}}$. If physical constraints exist, e.g.\ from the optical layout, that further limit their actually achievable range, such constraints can easily be implemented by restricting the range of values that the respective registers are allowed to assume, further increasing the robustness. In particular the ellipticity $\varepsilon$ will be almost constant in many cases, since it mainly depends on geometrical imperfections of the laser beams.

Each loop can now be individually optimized for the dynamical behaviour and noise characteristics of the corresponding signal. This concerns not only the servo loop bandwidths, but also the $2f$ filters, which can now be made more efficient for the $\alpha$, $\beta$ and $\varepsilon$ loops. 

\added{Going one step further, one could even think of using an optimized Kalman-type estimator for $\alpha$, $\beta$ and $\epsilon$ instead of a simple PI controller. The angles $\alpha$ and $\beta$ typically represent pitch and yaw of either the spacecraft or a testmass, and their dynamics consists of noise and commanded actuator signals. The latter are known and could even be used in a feedforward path to further reduce the burden on these filters.}

At first glance it may seem that this optical readout system could produce fundamentally different outputs, since the mixing matrix that implements Equations~(\ref{eq:ea})\dots (\ref{eq:ed}) acts on the error signals, which are weighted with the signal amplitudes in the individual segments A\dots D, whereas Equations~(\ref{eq:x})\dots (\ref{eq:eps}) act on the phases which have been stripped of the amplitude information. Further analysis shows, however, that this is not the case and that the proposed system produces the same outputs as the conventional one, at least if the servo loops have enough gain and integrator stages to keep all error signals close to zero. In that case, Equations~(\ref{eq:ea})\dots (\ref{eq:ed}) imply that the segment error signals ${E_A\dots E_D}$ are zero as well, which leads to the same outputs as in the conventional architecture.

\begin{figure}
	\centering
	\includegraphics{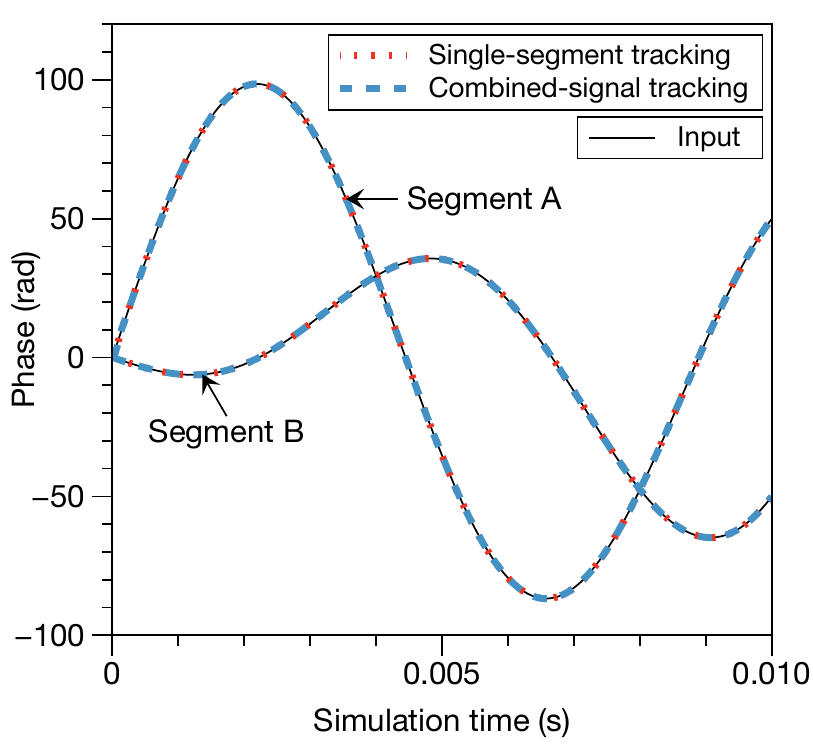}
	\caption{\added{Simulation of the conventional (single-segment tracking) and the proposed (combined-signal tracking) differential wavefront sensing architectures with DPLL readout, showing the successful tracking of two segments of a quadrant photodiode. Given the same input with no noise, and identical loop parameters, the two architectures produce the same output.}}
	\label{figure:simulink}
\end{figure}

\added{The proposed architecture has been verified to produce the same outputs as the conventional scheme by means of computer simulations and electrical tests~\cite{NILS}. For example, using a dynamical DPLL model implemented in MATLAB Simulink we show that, in the absence of noise, and with identical loop parameters, the conventional and the proposed architectures are mathematically equivalent (Figure~\ref{figure:simulink}).}

\added{Initial tests of the proposed architecture using optical signals have been performed in a simple Mach-Zehnder Interferometer. The light sources were two Nd:YAG NPRO lasers phase-locked to each other with 9\,MHz offset with separate electronics. The signals were recorded with an InGaAs QPD of 1\,mm diameter and were converted to voltages with op-amp based transimpedance amplifiers with 20\,MHz bandwidth. The phasemeter operated at 80\,MHz clock frequency and used the hardware which is described in~\cite[Ch.~9]{OLI}. Other than implementing the mixing matrices shown in Figure~\ref{fig:dwsnew}, the loop parameters were not changed between the conventional and the proposed schemes. The initial measurements confirmed that the scheme works functionally and is able to determine correct length and DWS signals. The length loop in the proposed scheme can indeed handle signals of roughly half the amplitude ($-6$\,dB) compared to the standard scheme, both in acquiring the signal and in tracking without cycle slipping, when the artificially increased noise floor remains unchanged, or with the same signal amplitude and 6\,dB more noise. Further experiments are being prepared using a LISA-representative testbed~\cite{Chwalla2020}.}

\section{Noise investigations}
\label{sec:tests}

\begin{figure}[b]
	\centering
	\includegraphics{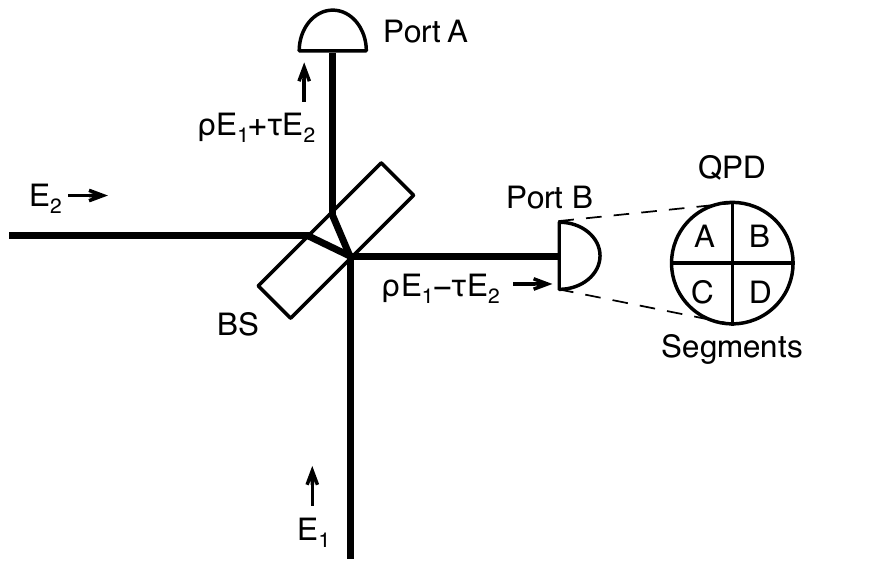}
	\caption{Diagram showing the interferometer being considered. $E_1$ and $E_2$ interfere at a beamsplitter (BS), and a quadrant photodiode is placed at each of the output ports.}
	\label{figure:setup}
\end{figure}

We analyze the proposed technique in the context of a heterodyne interferometer in the presence of relative intensity noise (RIN) from the laser sources, which is uncorrelated among the interfering laser beams, but correlated among QPD segments; and photon shot noise, which is uncorrelated among QPD segments. This analysis can be easily scaled to include any number of correlated and uncorrelated noise sources. Let $P_{jk}$ denote the optical power impinging in segment $k \in \{A,B,C,D\}$ of the photodiode located in port $j \in \{A,B\}$ of the interferometer consisting of a pair of QPDs mounted at the output ports of a beam combiner (BS), and let $E_1$ and $E_2$ denote the on-axis complex amplitudes of the interfering fields (Figure~\ref{figure:setup}). Upon combination, the superposed fields have complex amplitudes given by
\begin{align}
\mathrm{Port\,A:} \quad E_A &= \rho E_1 + \tau E_2, \\
\mathrm{Port\,B:} \quad E_B &= \rho E_1 + \tau e^{i\pi} E_2 = \rho E_1 - \tau E_2,
\end{align}
where $\rho$ and $\tau$ are the beamsplitter's amplitude reflection and transmission coefficients respectively. The photodiode signal in the quadrant labeled by $\left\{ j,k\right\}=\{\mathrm{port},\mathrm{segment}\}$ is given by $C_{jk} = c_{jk} P_{jk}$, where $c_{jk}$ is a constant determined by the analog processing of the signal. The optical power $P_{jk}$ is calculated as
\begin{equation}
P_{jk} = \iint \abs{E_j}^2 dS_{jk},
\end{equation}
where $dS_{jk}$ is a surface element in segment $k$ of the photodiode in port $j$. For example, for a segment in port $A$ we have
\begin{align}
P_{Ak} =& \rho^2 \iint \abs{E_1}^2 dS_{Ak} + \tau^2 \iint \abs{E_2}^2 dS_{Ak} \nonumber \\
&+ \rho \tau \left( \iint E_1 E_2^{\ast} dS_{Ak} +  \iint E_1^{\ast} E_2 dS_{Ak} \right).
	\label{equation:PAk}
\end{align}

Here we introduce the definition of the overlap integral as
\begin{equation}
\sqrt{\eta_{jk}}e^{i\phi_{jk}} \equiv \frac{\iint E_1 E_2^{\ast}dS_{jk}}{\left(\iint \abs{E_1}^2dS_{jk} \iint \abs{E_2}^2dS_{jk} \right)^{\frac{1}{2}}},
\end{equation}
where $\eta_{jk}$ is the heterodyne efficiency and $\phi_{jk} = 2\pi f_{\mathrm{het}} t - \psi_{jk}$ is the phase of the superposed optical fields, with $f_{\mathrm{het}}$ the heterodyne frequency (i.e., the beatnote frequency between $E_1$ and $E_2$), and $\psi_{jk}$ the interferometric phase. Recasting Equation~\ref{equation:PAk} in terms of $\eta_{Ak}$ and $\phi_{Ak}$ yields
\begin{align}
		P_{Ak} &= \frac{1}{4} \left( \rho^2 P_1 + \tau^2 P_2 + \rho \tau \sqrt{\eta_{Ak} P_1 P_2} \left[ e^{i\phi_{Ak}} + e^{-i \phi_{Ak}}  \right] \right) \nonumber \\
	&= \frac{1}{4} \left( \rho^2 P_1 + \tau^2 P_2 + 2\rho \tau \sqrt{\eta_{Ak} P_1 P_2} \cos \phi_{Ak}  \right),
	\label{equation:P_Ak}
\end{align}
where $P_1$ and $P_2$ represent the instantaneous power in the interfering beams, and we made the assumption that the beams are well-centered in the detectors such that the power is equally distributed over the four QPD segments. This is a reasonable assumption for well aligned interferometers, as well as for interferometers employing imaging systems to minimize beam walk at the detectors. RIN is a multiplicative noise source, and it can be modeled as
\begin{align}
P_{1} = \bar P_{1} + \delta P_{1} = \bar P_{1} (1+r_{1}), \\
P_{2} = \bar P_{2} + \delta P_{2} = \bar P_{2} (1+r_{2}),
\end{align}
where $\bar P_1$ and $\bar P_2$ represent the average power in the interfering fields, and RIN is defined as the power noise normalized to the average power, $r_i = \delta P_i / \bar P_i$ with $i=\left\{1,2\right\}$. By making the appropriate substitutions in Equation~\ref{equation:P_Ak}, we obtain
\begin{align}
P_{Ak} &= \frac{1}{4}  \rho^2 \bar P_1 (1+r_1) + \frac{1}{4} \tau^2 \bar P_2(1+r_2) \nonumber \\
&  + \frac{1}{2} \rho \tau \sqrt{\eta_{Ak} \bar P_1 \bar P_2 (1+r_1) (1+r_2)} \cos \phi_{Ak} , \nonumber \\
& \approx \frac{1}{4} \left( \rho^2 \bar P_1 + \tau^2 \bar P_2 + \rho^2 \bar P_1 r_1 + \tau^2 \bar P_2 r_2 \right)  \nonumber \\
& + \frac{1}{2} \rho \tau \left( 1 + \frac{r_1+r_2}{2} \right) \sqrt{\eta_{Ak} \bar P_1 \bar P_2 } \cos \phi_{Ak}
\end{align}
where we have used $\sqrt{1+r} \approx 1+r/2$, and we have neglected the small cross term $r_1r_2$. The term $\frac{1}{4} (\rho^2 \bar P_1 + \tau^2 \bar P_2)$ is the DC coupling, and the term $\frac{1}{4} (\rho^2 \bar P_1 r_1 + \tau^2 \bar P_2 r_2)$ is known as $1f$-RIN and is additive noise. The beatnote signal is mixed with $2f$-RIN proportional to $(r_1+r_2)/2$. \added{The $2f$-RIN coupling to the interferometric phase is independent of the beam powers and the heterodyne efficiency, and it is maximum for $\bar P_1 = \bar P_2$, where the $2f$-RIN coupling is $1/2$ that of the corresponding $1f$-RIN coupling. Since the total RIN coupling is the root sum square of the two contributions, neglecting $2f$-RIN induces only a small relative error of maximum $1-1/\sqrt{1+(1/2)^2} \approx 10\%$, which becomes much less significant if the beams have different power (e.g., it is less than $3\%$ for $\bar P_1 / \bar P_2 = 10$). To simplify our analysis, we disregard this small term}. The photodiode segment signals can then be written as the combination of three terms,
\begin{equation}
C_{jk} = c_{jk} P_{jk} = \DC_{jk} + r_{jk} + \beat_{jk},
\end{equation}
where $\DC_{jk}$ is the DC coupling of segment $k$ in port $j$, $r_{jk}$ is the $1f$-RIN coupling, and $b_{jk}$ is the beatnote signal. For example, for port $A$,
\begin{align}
\DC_{Ak} &= \frac{1}{4} c_{Ak} \left(\rho^2 \bar P_1 + \tau^2 \bar P_2 \right), \\
r_{Ak} &= \frac{1}{4} c_{Ak} \left( \rho^2 \bar P_1 r_1 + \tau^2 \bar P_2 r_2 \right), \\
\beat_{jk} &= \frac{1}{2} \rho \tau c_{jk} \sqrt{\eta_{jk} \bar P_1 \bar P_2 } \cos \phi_{jk}.
\end{align}
To characterize the resulting interferometric phase error, we calculate the inverse carrier-to-noise density
\begin{equation}
	\widetilde \psi_{jk} = \frac{\text{Noise ASD in $C_{jk}$}}{\text{Signal RMS in $C_{jk}$}} = \frac{ \widetilde {\Delta C}_{jk}}{\RMS{\beat_{jk}}},
	\label{equation:PN}
\end{equation}
where $\widetilde \psi_{jk}$ denotes the amplitude spectral density (ASD) of the interferometric phase, $\text{RMS}\{\beat_{jk}\}$ is the root mean square (RMS) of the beatnote signal, and $\widetilde{\Delta C}_{jk}$ is the ASD of the noise in the photodiode segment signal,
\begin{equation}
\widetilde{\Delta C}_{jk} = \sqrt{\tilde \sn_{jk}^2 + \tilde r_{jk}^2},
\end{equation}
where $\tilde \sn_{jk}$ is the ASD of the shot noise coupling, and $\tilde r_{jk}$ is the ASD of the $1f$-RIN coupling. Shot noise of the photodetection is white noise affecting all frequencies, and typically it is the limiting noise source at high frequency. The ASD of shot noise is given for each detector segment as
\begin{align}
\tilde \sn_{jk} &= \sqrt{2 q_e \DC_{jk} },
\end{align}
where $q_e$ is the electron charge. Shot noise adds quadratically with $1f$-RIN to the total phase noise. The $1f$-RIN coupling ASD for a segment in port $A$ is given by
\begin{equation}
	\tilde r_{Ak} = \frac{1}{4} c_{Ak} \left( \rho^2 \bar P_1 \tilde r_1 + \tau^2 \bar P_2 \tilde r_2 \right).
	\label{equation:1f-rin}
\end{equation}
The addition in Equation~\ref{equation:1f-rin} is performed linearly if the RIN in $E_1$ and $E_2$ is correlated, and quadratically otherwise. The RMS value of the signal is
\begin{equation}
	\RMS{\beat_{jk}} = \frac{1}{4} \rho \tau c_{jk} \sqrt{2 \eta_{jk} \bar P_1 \bar P_2},
\end{equation}
where we have used the definition of the RMS value of a function of period $T$,
\begin{equation}
	\RMS{f(t)} = \sqrt{\frac{1}{T} \int_{0}^{T} \left[ f(t) \right]^2 dt}.
\end{equation}
The phase noise for a segment in port A considering $1f$-RIN coupling only is given by
\begin{equation}
	\widetilde \psi_{Ak} = \frac{  \rho^2 \bar P_1 \tilde r_1 + \tau^2 \bar P_2 \tilde r_2  }{\rho \tau \sqrt{2 \eta_{Ak} \bar P_1 \bar P_2}}.
	\label{equation:1f-rin-segment}
\end{equation}
The phase noise due to shot noise only is given by
\begin{equation}
	\widetilde \psi_{Ak} = \frac{2}{\rho \tau} \sqrt{ \frac{q_e  \left( \rho^2 \bar P_1 + \tau^2 \bar P_2\right) }{  c_{Ak} \eta_{Ak} \bar P_1 \bar P_2}}.
	\label{equation:shot-noise-segment}
\end{equation}
Each QPD segment ``sees'' $1/4$ of the signal, and $1/4$ of the $1f$-RIN, and hence has the same phase error associated with $1f$-RIN coupling as a single element photodiode (SEPD) with four times the surface area. However, each segment sees up to $1/2$ of the shot noise as the SEPD, resulting in up to twice the phase error associated with shot noise, or with any source of noise that is uncorrelated among segments, as the SEPD. The total phase noise in port $A$ of the interferometer is given by
\begin{equation}
\widetilde \psi_{Ak} = \frac{\sqrt{\tilde \sn_{Ak}^2 + \tilde r_{Ak}^2 }} { \frac{1}{4} \rho \tau c_{Ak} \sqrt{2 \eta_{Ak} \bar P_1 \bar P_2}},
\label{equation:total-phase-noise-QPDsegment}
\end{equation}
where we have assumed that the RIN noise in $E_1$ and $E_2$ is uncorrelated, so that $\tilde r_1$ and $\tilde r_2$ add quadratically. In the conventional DWS architecture, the PLL acts on the signals $C_{jk}$ of the individual QPD segments, and thus the performance is limited by the phase noise given in Equation~\ref{equation:total-phase-noise-QPDsegment}. In the proposed DWS architecture, the $C_{jk}$ signals are first combined linearly, and the PLL acts on the linear combinations. The shot noise is uncorrelated among segments and adds quadratically, whereas $1f$-RIN is correlated and adds linearly.
 
For example, for tracking the length signal, we combine the four QPD quadrants in port $A$ as 
\begin{equation}
	C_{A(A+B+C+D)}=\sumqpd C_{Ak} = \sumqpd \DC_{Ak} + \sumqpd r_{Ak} + \sumqpd \beat_{Ak}.
\end{equation}
We omit the factor $1/4$ in the signal combination (see Equation~\ref{eq:ea}), since it affects equally the noise ASD and the beatnote signal RMS of the combined signal, and therefore has no impact on the SNR of the tracked signal. Note that in this case the PLL does not track the phase of the beatnote signal $E_A = \rho E_1 + \tau E_2$, but instead tracks the phase of the beatnote signal that results from adding the four QPD segments $A$, $B$, $C$, and $D$. The noise ASD of the combined signal is given by
\begin{equation}
	\widetilde {\Delta C}_{A(A+B+C+D)} = \left(\sumqpd \tilde \sn_{Ak}^2 + \left[ \sumqpd \tilde r_{Ak} \right]^2 \right)^{\frac{1}{2}},
	\label{equation:noise-ASD-4Q}
\end{equation}
and the RMS value of the signal is given by
\begin{widetext}
\begin{equation}
\RMS{\sumqpd \beat_{Ak}} = \frac{\rho \tau \sqrt{\bar P_1 \bar P_2}}{2 \sqrt{2}} \left( \sum\limits_{k_1,k_2=A}^{D} c_{Ak_1}c_{Ak_2} \sqrt{\eta_{Ak_1}\eta_{Ak_2}} \right)^{\frac{1}{2}} = \frac{\rho \tau \sqrt{\bar P_1 \bar P_2}}{2 \sqrt{2}} \left( \sumqpd c_{Ak}^2 \eta_{Ak} + 2\!\!\!\!\!\! \sum\limits_{A\leq k_1 < k_2 \leq D} \!\!\!\! c_{Ak_1}c_{Ak_2}\sqrt{\eta_{Ak_1}\eta_{Ak_2}} \right)^{\frac{1}{2}}.
\label{equation:RMS-4Q}
\end{equation}
\end{widetext}
where $k_1,k_2 \in \left\{A,B,C,D\right\}$. Equations~\ref{equation:noise-ASD-4Q} and~\ref{equation:RMS-4Q} can be substituted into Equation~\ref{equation:PN} to obtain the total noise of the combined signal of the four QPD segments. The resulting expression is quite long. By assuming identical analog signal processing $c_{Ak_1} = c_{Ak_2} = c_{A}$, and segment heterodyne efficiencies $\eta_{Ak_1} = \eta_{Ak_2} = \eta_A$, we obtain the much simpler expression
\begin{equation}
	\widetilde E_{A(A+B+C+D)} \approx \frac{1}{\rho\tau} \sqrt{ \frac{2 q_e (\rho^2 \bar P_1 + \tau^2 \bar P_2) + c_A(\rho^4 \bar P_1^2 \tilde r_1^2 + \tau^4 \bar P_2 \tilde r_2^2) }{2 c_A \eta_A \bar P_1 \bar P_2 } }.
	\label{equation:total-noise-4Q}
\end{equation}
The combined signal has the same level of $1f$-RIN coupling as in the single-segment case (Equation~\ref{equation:1f-rin-segment}), but lower shot noise coupling. Shot noise in Equation~\ref{equation:total-noise-4Q} is $1/2$ that of the single-segment case (Equation~\ref{equation:shot-noise-segment}).

\begin{figure}
	\centering
	\includegraphics{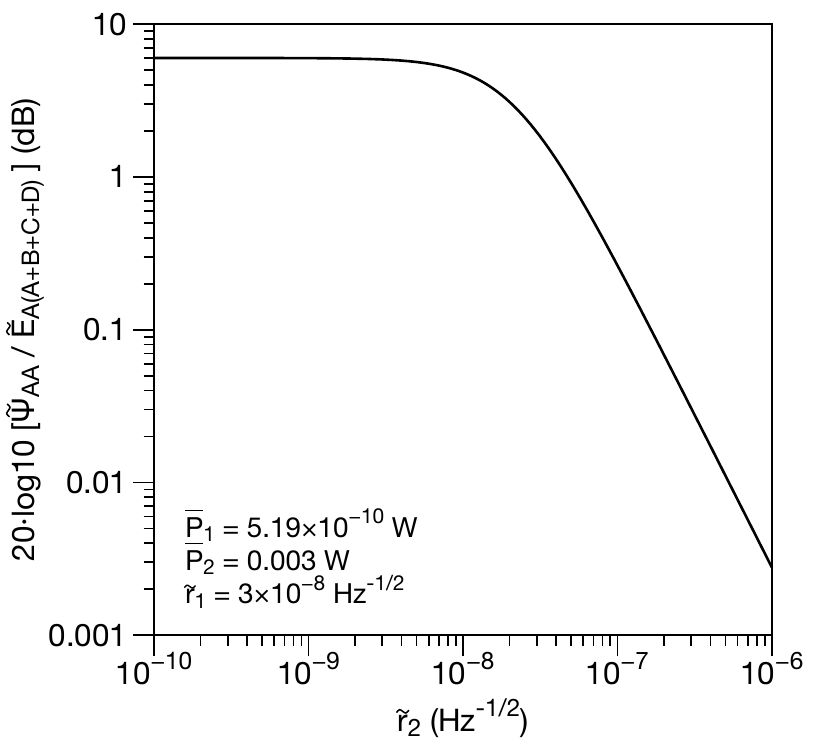}
	\caption{Improvement of the proposed scheme over single-segment tracking as a function of the RIN level in the strong field, for an interferometer resembling the LISA long-arm interferometer.}
	\label{figure:improvement}
\end{figure}

Equation~\ref{equation:total-phase-noise-QPDsegment} (i.e., the phase noise in single-segment tracking) and Equation~\ref{equation:total-noise-4Q} can be compared to assess the improvement of the proposed technique:
\begin{equation}
	\frac{\widetilde E_{A(A+B+C+D)}}{\widetilde \psi_{AA}} \approx \sqrt{1-\frac{12 q_e \left( \bar P_1 + \bar P_2 \right)}{16 q_e \left(\bar P_1 + \bar P_2 \right) + c_A \left( \bar P_1^2 \tilde r_1^2 + \bar P_2^2 \tilde r_2^2 \right)  }}.
\end{equation}
It can now be readily seen that the proposed technique provides a maximum improvement of 6\,dB over the traditional scheme,
\begin{equation}
	20\cdot \log \left( \frac{\widetilde E_{A(A+B+C+D)}}{\widetilde \psi_{AA}} \right) \approx 20 \cdot \log \left( \sqrt{ 1-\frac{3}{4}} \right) \approx -6\,\text{dB}.
\end{equation}
When tracking the interferometric phase signal from a single segment of a QPD, the PLL ``sees'' $1/4$ of the signal RMS, and $1/4$ of the $1f$-RIN, but up to $1/2$ of the shot noise, as a SEPD. This results in up to twice the shot noise induced phase error as when tracking the phase signal of the SEPD. To take full advantage of the QPD, in the proposed architecture the PLL is applied not to the individual QPD segments, but to the linear combinations of these signals (Equations~\ref{eq:ea}-\ref{eq:ed}). In this way, the tracked signals will have the same level of correlated noise coupling such as $1f$-RIN as in single-segment tracking, but up to $2$ times less uncorrelated noise coupling, such as shot noise, resulting in an overall similar noise performance as when tracking the signal from a SEPD of the same size as the QPD. This is in addition to other advantages such as being able to adjust the loop gains and filter parameters individually for length and angular signals.

In the LISA long-arm interferometer, one field is much stronger than the other. In this case we may simplify further (i.e., $\bar P_2 \gg \bar P_1,\,\bar P_2^2 \tilde r_2^2 \gg \bar P_1^2 \tilde r_1^2$),
\begin{equation}
	\frac{\widetilde E_{A(A+B+C+D)}}{\widetilde \psi_{AA}} \approx \sqrt{1-\frac{12 q_e \bar P_2 }{16 q_e \bar P_2 + c_a \bar P_2^2 \tilde r_2^2}}.
\end{equation}
Figure~\ref{figure:improvement} shows the improvement as a function of the RIN level in the strong field in an interferometer with parameters akin to the LISA long-arm interferometer.

A major advantage of the increased SNR in the proposed architecture is the enhanced robustness against alignment errors of the interferometer. This is an issue of great concern in the development of long-baseline space-based observatories, such as LISA, where the unavoidable angular motion of the spacecraft translates into alignment jitter in the interferometer, and therefore in a degradation of heterodyne efficiency and a cross-coupling in the longitudinal pathlength measurements known as tilt-to-length (TTL) coupling. Imaging systems can be used to reduce TTL-coupling in the interferometer to a great extent~\cite{Trobs2018, Chwalla2020}. The imaging systems suppress beam walk in the detector plane, thus leading to the reduction in TTL-coupling, but in turn they magnify the beam tilt angle. The angular magnification provided by the imaging systems, in conjunction with the one provided by the receiving telescope, means that the phasemeter should be able to track signals with relatively poor heterodyne efficiency.

\begin{figure}
	\centering
	\includegraphics{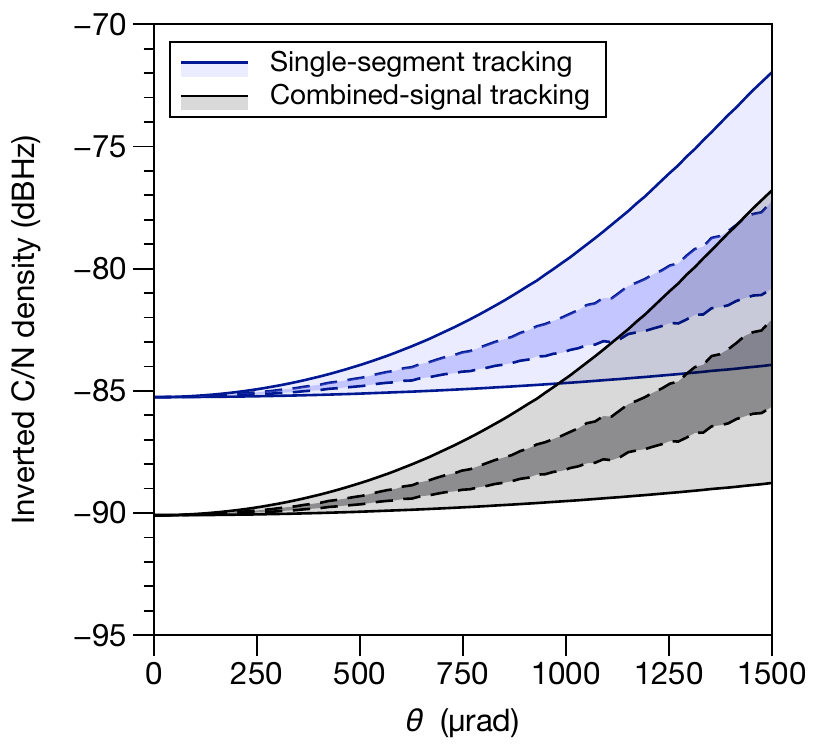}
	\caption{Inverted carrier-to-noise density of the signals being tracked by the PLL in the conventional (single-segment tracking) and the proposed (combined-signal tracking) architectures, resulting from a Monte-Carlo simulation of a misaligned interferometer where the beam tilt angle, as sensed by the detector, is sampled from a uniform distribution with amplitude $\theta$, and the beam tilt direction is sampled from a uniform distribution spanning the entire plane transverse to the optical axis. The beam tilts with respect to the center of the detector as if it were being imaged to that point by an imaging system. In single-segment tracking, all four QPD segments are taken into consideration in the statistical analysis. The solid lines represent the maxima and minima of the resulting inverted C/N density distributions. The dark-shade regions delimited by dashed lines represent one standard deviation from the mean.}
	\label{figure:improvement-2}
\end{figure}

\begin{table}
\caption{Simulation parameters. The model assumes identical analog processing of the QPD segments.}
\label{thetable}
\centering
\begin{tabular}{lr}
\toprule
Property & Value \\
\midrule
$\bar P_1$ & 519\,pW \\
$\bar P_2$ & 3\,mW \\
$\tilde r_1$ & $10^{-8}/\sqrt{\mathrm{Hz}}$ \\
$\tilde r_2$ & $10^{-8}/\sqrt{\mathrm{Hz}}$ \\
$\rho$ = $\tau$ & $1/\sqrt{2}$ \\
QPD active area diameter & 1\,mm \\
QPD slit width & 20\,$\mu$m \\
Reference beam waist radius & 0.5\,mm \\
Measurement beam waist radius & 10\,mm \\
\bottomrule
\end{tabular}
\end{table}

To quantify the improvement of the proposed technique under alignment errors of the interferometer, we develop a computer model of the interferometer using Ifocad. Ifocad is a collection of C++ libraries that provides proven methods for simulating two-beam laser interferometers~\cite{Ifocad, DWSFAC}. We simulate the interference between a Gaussian reference beam with typical parameters (LO beam), and a Gaussian beam with a very large waist located at the position of a QPD such that, at this position, it resembles the flat-top beam received by a LISA spacecraft (Rx beam). This particular simulation method has been verified experimentally in a testbed featuring flat-top beam on Gaussian beam interference~\cite{Chwalla2020}. 

We perform a Monte-Carlo analysis in which the program computes the heterodyne efficiencies for each of the four segments of the QPD, as the Rx beam is rotated with respect to the center of the detector, so as to mimic the effect of an imaging system that has been aligned perfectly and is therefore suppressing the Rx beam walk in the detector plane. The beam	tilt angle is sampled from a uniform distribution with increasing amplitude, and the tilt direction is sampled from a uniform distribution spanning the entire plane transverse to the optical axis.	The program then uses the computed heterodyne efficiencies to calculate the estimated signal error in the conventional (single-segment tracking, Equation~\ref{equation:total-phase-noise-QPDsegment}) and the proposed (combined-signal tracking, Equations~\ref{equation:RMS-4Q} and~\ref{equation:noise-ASD-4Q}) architectures. The simulation parameters are gathered in Table~\ref{thetable}. 

As expected, the proposed technique offers improved SNR throughout the tilt range (Figure~\ref{figure:improvement-2}). It is therefore able to handle worse absolute beam misalignments and higher residual beam pointing jitter. This improvement is crucial for a enhanced tolerance of spacecraft attitude control instabilities prior to acquiring a stable phase locking of a slave laser to a weak received beam, as required for the establishment of an inter-spacecraft laser link in a laser transponder configuration. This improved SNR is then specially relevant throughout the initial beatnote acquisition states and re-acquisition sequences~\cite{Wuchenich2014, Koch2020}, providing more robust control loops for sensing the optical link dynamics.

It is important to note that the noise improvement is achieved in the signals being tracked by each separate PLL instance, and not on the length and angular signals. In the conventional scheme, the individual QPD segment signals tracked by four PLLs are combined in the next stage to yield length and angular signals with the same level of noise as in the proposed scheme. The advantage of the proposed architecture lies in the enhancement of the tracking loops: the PLLs track signals with greater SNR, which results in more robust locks with lower probability of cycle slipping, and the added benefit of being able to tune each PLL instance to optimally track its corresponding length or angular signal channel.

\section{Conclusion}
\label{sec:conclusion}

We have proposed a scheme to process the heterodyne beatnotes from the four segments of a quadrant photodiode in heterodyne interferometers that use DPLL-based phasemeters. It acts on the length signal $x$ and angular signals $\alpha$, $\beta$ and $\varepsilon$, which directly correspond to physically meaningful parameters.
The proposed scheme has advantages in terms of robustness against cycle slips. It allows to individually optimize the loop gains and filter parameters for length and angular signals, which may also lead to lower noise in these outputs. 
A noise analysis of the proposed architecture has been carried out in the context of a heterodyne interferometer in the presence of shot noise and relative intensity noise. Then, a computer model was used to analyze the degradation of the tracked signals in a misaligned interferometer. The proposed technique has been compared against the conventional one to demonstrate a maximum improvement of 6\,dB.
We expect that this scheme may be an attractive alternative for applications like LISA or future GRACE Follow-On-like geodesy missions. \added{It can also find applications across multiple disciplines where high precision length and angular measurements are desired, notably in inertial sensing of test masses, with applications in fields such as vibration isolation~\cite{MowLowry2019}, and tests of fundamental physics~\cite{Schlamminger2008, Luo2009, Li2018}.}

\section{Acknowledgements}

The authors acknowledge the support of the German Space Agency, DLR (FKZ 50OQ1801); Clusters of Excellence ``QuantumFrontiers: Light and Matter at the Quantum Frontier: Foundations and Applications in Metrology'' (EXC-2123, project number: 390837967); PhoenixD: ``Photonics, Optics, and Engineering—Innovation Across Disciplines'' (EXC-2122, project number: 390833453).

\section{References}

\begin{thebibliography}{35}%
\makeatletter
\providecommand \@ifxundefined [1]{%
 \@ifx{#1\undefined}
}%
\providecommand \@ifnum [1]{%
 \ifnum #1\expandafter \@firstoftwo
 \else \expandafter \@secondoftwo
 \fi
}%
\providecommand \@ifx [1]{%
 \ifx #1\expandafter \@firstoftwo
 \else \expandafter \@secondoftwo
 \fi
}%
\providecommand \natexlab [1]{#1}%
\providecommand \enquote  [1]{``#1''}%
\providecommand \bibnamefont  [1]{#1}%
\providecommand \bibfnamefont [1]{#1}%
\providecommand \citenamefont [1]{#1}%
\providecommand \href@noop [0]{\@secondoftwo}%
\providecommand \href [0]{\begingroup \@sanitize@url \@href}%
\providecommand \@href[1]{\@@startlink{#1}\@@href}%
\providecommand \@@href[1]{\endgroup#1\@@endlink}%
\providecommand \@sanitize@url [0]{\catcode `\\12\catcode `\$12\catcode
  `\&12\catcode `\#12\catcode `\^12\catcode `\_12\catcode `\%12\relax}%
\providecommand \@@startlink[1]{}%
\providecommand \@@endlink[0]{}%
\providecommand \url  [0]{\begingroup\@sanitize@url \@url }%
\providecommand \@url [1]{\endgroup\@href {#1}{\urlprefix }}%
\providecommand \urlprefix  [0]{URL }%
\providecommand \Eprint [0]{\href }%
\providecommand \doibase [0]{http://dx.doi.org/}%
\providecommand \selectlanguage [0]{\@gobble}%
\providecommand \bibinfo  [0]{\@secondoftwo}%
\providecommand \bibfield  [0]{\@secondoftwo}%
\providecommand \translation [1]{[#1]}%
\providecommand \BibitemOpen [0]{}%
\providecommand \bibitemStop [0]{}%
\providecommand \bibitemNoStop [0]{.\EOS\space}%
\providecommand \EOS [0]{\spacefactor3000\relax}%
\providecommand \BibitemShut  [1]{\csname bibitem#1\endcsname}%
\let\auto@bib@innerbib\@empty
\bibitem [{\citenamefont {Armano}\ \emph {et~al.}()\citenamefont {Armano},
  \citenamefont {Audley}, \citenamefont {Baird}, \citenamefont {Binetruy},
  \citenamefont {Born}, \citenamefont {Bortoluzzi}, \citenamefont {Castelli},
  \citenamefont {Cavalleri}, \citenamefont {Cesarini}, \citenamefont {Cruise},
  \citenamefont {Danzmann}, \citenamefont {de~Deus~Silva}, \citenamefont
  {Diepholz}, \citenamefont {Dixon}, \citenamefont {Dolesi}, \citenamefont
  {Ferraioli}, \citenamefont {Ferroni}, \citenamefont {Fitzsimons},
  \citenamefont {Freschi}, \citenamefont {Gesa}, \citenamefont {Gibert},
  \citenamefont {Giardini}, \citenamefont {Giusteri}, \citenamefont {Grimani},
  \citenamefont {Grzymisch}, \citenamefont {Harrison}, \citenamefont {Heinzel},
  \citenamefont {Hewitson}, \citenamefont {Hollington}, \citenamefont
  {Hoyland}, \citenamefont {Hueller}, \citenamefont {Inchauspe}, \citenamefont
  {Jennrich}, \citenamefont {Jetzer}, \citenamefont {Karnesis}, \citenamefont
  {Kaune}, \citenamefont {Korsakova}, \citenamefont {Killow}, \citenamefont
  {Lobo}, \citenamefont {Lloro}, \citenamefont {Liu}, \citenamefont
  {Lopez-Zaragoza}, \citenamefont {Maarschalkerweerd}, \citenamefont {Mance},
  \citenamefont {Meshksar}, \citenamefont {Martin}, \citenamefont
  {Martin-Polo}, \citenamefont {Martino}, \citenamefont {Martin-Porqueras},
  \citenamefont {Mateos}, \citenamefont {McNamara}, \citenamefont {Mendes},
  \citenamefont {Mendes}, \citenamefont {Nofrarias}, \citenamefont
  {Paczkowski}, \citenamefont {Perreur-Lloyd}, \citenamefont {Petiteau},
  \citenamefont {Pivato}, \citenamefont {Plagnol}, \citenamefont
  {Ramos-Castro}, \citenamefont {Reiche}, \citenamefont {Robertson},
  \citenamefont {Rivas}, \citenamefont {Russano}, \citenamefont {Slutsky},
  \citenamefont {Sopuerta}, \citenamefont {Sumner}, \citenamefont {Texier},
  \citenamefont {Thorpe}, \citenamefont {Vetrugno}, \citenamefont {Vitale},
  \citenamefont {Wanner}, \citenamefont {Ward}, \citenamefont {Wass},
  \citenamefont {Weber}, \citenamefont {Wissel}, \citenamefont {Wittchen},\
  and\ \citenamefont {Zweifel}}]{LTP2}%
  \BibitemOpen
  \bibfield  {author} {\bibinfo {author} {\bibfnamefont {M.}~\bibnamefont
  {Armano}}, \bibinfo {author} {\bibfnamefont {H.}~\bibnamefont {Audley}},
  \bibinfo {author} {\bibfnamefont {J.}~\bibnamefont {Baird}}, \bibinfo
  {author} {\bibfnamefont {P.}~\bibnamefont {Binetruy}}, \bibinfo {author}
  {\bibfnamefont {M.}~\bibnamefont {Born}}, \bibinfo {author} {\bibfnamefont
  {D.}~\bibnamefont {Bortoluzzi}}, \bibinfo {author} {\bibfnamefont
  {E.}~\bibnamefont {Castelli}}, \bibinfo {author} {\bibfnamefont
  {A.}~\bibnamefont {Cavalleri}}, \bibinfo {author} {\bibfnamefont
  {A.}~\bibnamefont {Cesarini}}, \bibinfo {author} {\bibfnamefont {A.~M.}\
  \bibnamefont {Cruise}}, \bibinfo {author} {\bibfnamefont {K.}~\bibnamefont
  {Danzmann}}, \bibinfo {author} {\bibfnamefont {M.}~\bibnamefont
  {de~Deus~Silva}}, \bibinfo {author} {\bibfnamefont {I.}~\bibnamefont
  {Diepholz}}, \bibinfo {author} {\bibfnamefont {G.}~\bibnamefont {Dixon}},
  \bibinfo {author} {\bibfnamefont {R.}~\bibnamefont {Dolesi}}, \bibinfo
  {author} {\bibfnamefont {L.}~\bibnamefont {Ferraioli}}, \bibinfo {author}
  {\bibfnamefont {V.}~\bibnamefont {Ferroni}}, \bibinfo {author} {\bibfnamefont
  {E.~D.}\ \bibnamefont {Fitzsimons}}, \bibinfo {author} {\bibfnamefont
  {M.}~\bibnamefont {Freschi}}, \bibinfo {author} {\bibfnamefont
  {L.}~\bibnamefont {Gesa}}, \bibinfo {author} {\bibfnamefont {F.}~\bibnamefont
  {Gibert}}, \bibinfo {author} {\bibfnamefont {D.}~\bibnamefont {Giardini}},
  \bibinfo {author} {\bibfnamefont {R.}~\bibnamefont {Giusteri}}, \bibinfo
  {author} {\bibfnamefont {C.}~\bibnamefont {Grimani}}, \bibinfo {author}
  {\bibfnamefont {J.}~\bibnamefont {Grzymisch}}, \bibinfo {author}
  {\bibfnamefont {I.}~\bibnamefont {Harrison}}, \bibinfo {author}
  {\bibfnamefont {G.}~\bibnamefont {Heinzel}}, \bibinfo {author} {\bibfnamefont
  {M.}~\bibnamefont {Hewitson}}, \bibinfo {author} {\bibfnamefont
  {D.}~\bibnamefont {Hollington}}, \bibinfo {author} {\bibfnamefont
  {D.}~\bibnamefont {Hoyland}}, \bibinfo {author} {\bibfnamefont
  {M.}~\bibnamefont {Hueller}}, \bibinfo {author} {\bibfnamefont
  {H.}~\bibnamefont {Inchauspe}}, \bibinfo {author} {\bibfnamefont
  {O.}~\bibnamefont {Jennrich}}, \bibinfo {author} {\bibfnamefont
  {P.}~\bibnamefont {Jetzer}}, \bibinfo {author} {\bibfnamefont
  {N.}~\bibnamefont {Karnesis}}, \bibinfo {author} {\bibfnamefont
  {B.}~\bibnamefont {Kaune}}, \bibinfo {author} {\bibfnamefont
  {N.}~\bibnamefont {Korsakova}}, \bibinfo {author} {\bibfnamefont {C.~J.}\
  \bibnamefont {Killow}}, \bibinfo {author} {\bibfnamefont {J.~A.}\
  \bibnamefont {Lobo}}, \bibinfo {author} {\bibfnamefont {I.}~\bibnamefont
  {Lloro}}, \bibinfo {author} {\bibfnamefont {L.}~\bibnamefont {Liu}}, \bibinfo
  {author} {\bibfnamefont {J.~P.}\ \bibnamefont {Lopez-Zaragoza}}, \bibinfo
  {author} {\bibfnamefont {R.}~\bibnamefont {Maarschalkerweerd}}, \bibinfo
  {author} {\bibfnamefont {D.}~\bibnamefont {Mance}}, \bibinfo {author}
  {\bibfnamefont {N.}~\bibnamefont {Meshksar}}, \bibinfo {author}
  {\bibfnamefont {V.}~\bibnamefont {Martin}}, \bibinfo {author} {\bibfnamefont
  {L.}~\bibnamefont {Martin-Polo}}, \bibinfo {author} {\bibfnamefont
  {J.}~\bibnamefont {Martino}}, \bibinfo {author} {\bibfnamefont
  {F.}~\bibnamefont {Martin-Porqueras}}, \bibinfo {author} {\bibfnamefont
  {I.}~\bibnamefont {Mateos}}, \bibinfo {author} {\bibfnamefont {P.~W.}\
  \bibnamefont {McNamara}}, \bibinfo {author} {\bibfnamefont {J.}~\bibnamefont
  {Mendes}}, \bibinfo {author} {\bibfnamefont {L.}~\bibnamefont {Mendes}},
  \bibinfo {author} {\bibfnamefont {M.}~\bibnamefont {Nofrarias}}, \bibinfo
  {author} {\bibfnamefont {S.}~\bibnamefont {Paczkowski}}, \bibinfo {author}
  {\bibfnamefont {M.}~\bibnamefont {Perreur-Lloyd}}, \bibinfo {author}
  {\bibfnamefont {A.}~\bibnamefont {Petiteau}}, \bibinfo {author}
  {\bibfnamefont {P.}~\bibnamefont {Pivato}}, \bibinfo {author} {\bibfnamefont
  {E.}~\bibnamefont {Plagnol}}, \bibinfo {author} {\bibfnamefont
  {J.}~\bibnamefont {Ramos-Castro}}, \bibinfo {author} {\bibfnamefont
  {J.}~\bibnamefont {Reiche}}, \bibinfo {author} {\bibfnamefont {D.~I.}\
  \bibnamefont {Robertson}}, \bibinfo {author} {\bibfnamefont {F.}~\bibnamefont
  {Rivas}}, \bibinfo {author} {\bibfnamefont {G.}~\bibnamefont {Russano}},
  \bibinfo {author} {\bibfnamefont {J.}~\bibnamefont {Slutsky}}, \bibinfo
  {author} {\bibfnamefont {C.~F.}\ \bibnamefont {Sopuerta}}, \bibinfo {author}
  {\bibfnamefont {T.}~\bibnamefont {Sumner}}, \bibinfo {author} {\bibfnamefont
  {D.}~\bibnamefont {Texier}}, \bibinfo {author} {\bibfnamefont {J.~I.}\
  \bibnamefont {Thorpe}}, \bibinfo {author} {\bibfnamefont {D.}~\bibnamefont
  {Vetrugno}}, \bibinfo {author} {\bibfnamefont {S.}~\bibnamefont {Vitale}},
  \bibinfo {author} {\bibfnamefont {G.}~\bibnamefont {Wanner}}, \bibinfo
  {author} {\bibfnamefont {H.}~\bibnamefont {Ward}}, \bibinfo {author}
  {\bibfnamefont {P.~J.}\ \bibnamefont {Wass}}, \bibinfo {author}
  {\bibfnamefont {W.~J.}\ \bibnamefont {Weber}}, \bibinfo {author}
  {\bibfnamefont {L.}~\bibnamefont {Wissel}}, \bibinfo {author} {\bibfnamefont
  {A.}~\bibnamefont {Wittchen}}, \ and\ \bibinfo {author} {\bibfnamefont
  {P.}~\bibnamefont {Zweifel}},\ }\bibfield  {title} {\enquote {\bibinfo
  {title} {Beyond the required lisa free-fall performance: New lisa pathfinder
  results down to $20\,\mu$hz},}\ }\href {\doibase
  10.1103/PhysRevLett.120.061101} {\bibfield  {journal} {\bibinfo  {journal}
  {Physical Review Letters}\ }\textbf {\bibinfo {volume} {120}},\
  10.1103/PhysRevLett.120.061101}\BibitemShut {NoStop}%
\bibitem [{\citenamefont {Abich}\ \emph {et~al.}(2019)\citenamefont {Abich},
  \citenamefont {Abramovici}, \citenamefont {Amparan}, \citenamefont
  {Baatzsch}, \citenamefont {Okihiro}, \citenamefont {Barr}, \citenamefont
  {Bize}, \citenamefont {Bogan}, \citenamefont {Braxmaier}, \citenamefont
  {Burke}, \citenamefont {Clark}, \citenamefont {Dahl}, \citenamefont {Dahl},
  \citenamefont {Danzmann}, \citenamefont {Davis}, \citenamefont {de~Vine},
  \citenamefont {Dickson}, \citenamefont {Dubovitsky}, \citenamefont {Eckardt},
  \citenamefont {Ester}, \citenamefont {Barranco}, \citenamefont {Flatscher},
  \citenamefont {Flechtner}, \citenamefont {Folkner}, \citenamefont {Francis},
  \citenamefont {Gilbert}, \citenamefont {Gilles}, \citenamefont {Gohlke},
  \citenamefont {Grossard}, \citenamefont {Guenther}, \citenamefont {Hager},
  \citenamefont {Hauden}, \citenamefont {Heine}, \citenamefont {Heinzel},
  \citenamefont {Herding}, \citenamefont {Hinz}, \citenamefont {Howell},
  \citenamefont {Katsumura}, \citenamefont {Kaufer}, \citenamefont {Klipstein},
  \citenamefont {Koch}, \citenamefont {Kruger}, \citenamefont {Larsen},
  \citenamefont {Lebeda}, \citenamefont {Lebeda}, \citenamefont {Leikert},
  \citenamefont {Liebe}, \citenamefont {Liu}, \citenamefont {Lobmeyer},
  \citenamefont {Mahrdt}, \citenamefont {Mangoldt}, \citenamefont {McKenzie},
  \citenamefont {Misfeldt}, \citenamefont {Morton}, \citenamefont {M\"uller},
  \citenamefont {Murray}, \citenamefont {Nguyen}, \citenamefont {Nicklaus},
  \citenamefont {Pierce}, \citenamefont {Ravich}, \citenamefont {Reavis},
  \citenamefont {Reiche}, \citenamefont {Sanjuan}, \citenamefont {Sch\"utze},
  \citenamefont {Seiter}, \citenamefont {Shaddock}, \citenamefont {Sheard},
  \citenamefont {Sileo}, \citenamefont {Spero}, \citenamefont {Spiers},
  \citenamefont {Stede}, \citenamefont {Stephens}, \citenamefont {Sutton},
  \citenamefont {Trinh}, \citenamefont {Voss}, \citenamefont {Wang},
  \citenamefont {Wang}, \citenamefont {Ware}, \citenamefont {Wegener},
  \citenamefont {Windisch}, \citenamefont {Woodruff}, \citenamefont {Zender},\
  and\ \citenamefont {Zimmermann}}]{PhysRevLett.123.031101}%
  \BibitemOpen
  \bibfield  {author} {\bibinfo {author} {\bibfnamefont {Klaus}\ \bibnamefont
  {Abich}}, \bibinfo {author} {\bibfnamefont {Alexander}\ \bibnamefont
  {Abramovici}}, \bibinfo {author} {\bibfnamefont {Bengie}\ \bibnamefont
  {Amparan}}, \bibinfo {author} {\bibfnamefont {Andreas}\ \bibnamefont
  {Baatzsch}}, \bibinfo {author} {\bibfnamefont {Brian~Bachman}\ \bibnamefont
  {Okihiro}}, \bibinfo {author} {\bibfnamefont {David~C.}\ \bibnamefont
  {Barr}}, \bibinfo {author} {\bibfnamefont {Maxime~P.}\ \bibnamefont {Bize}},
  \bibinfo {author} {\bibfnamefont {Christina}\ \bibnamefont {Bogan}}, \bibinfo
  {author} {\bibfnamefont {Claus}\ \bibnamefont {Braxmaier}}, \bibinfo {author}
  {\bibfnamefont {Michael~J.}\ \bibnamefont {Burke}}, \bibinfo {author}
  {\bibfnamefont {Ken~C.}\ \bibnamefont {Clark}}, \bibinfo {author}
  {\bibfnamefont {Christian}\ \bibnamefont {Dahl}}, \bibinfo {author}
  {\bibfnamefont {Katrin}\ \bibnamefont {Dahl}}, \bibinfo {author}
  {\bibfnamefont {Karsten}\ \bibnamefont {Danzmann}}, \bibinfo {author}
  {\bibfnamefont {Mike~A.}\ \bibnamefont {Davis}}, \bibinfo {author}
  {\bibfnamefont {Glenn}\ \bibnamefont {de~Vine}}, \bibinfo {author}
  {\bibfnamefont {Jeffrey~A.}\ \bibnamefont {Dickson}}, \bibinfo {author}
  {\bibfnamefont {Serge}\ \bibnamefont {Dubovitsky}}, \bibinfo {author}
  {\bibfnamefont {Andreas}\ \bibnamefont {Eckardt}}, \bibinfo {author}
  {\bibfnamefont {Thomas}\ \bibnamefont {Ester}}, \bibinfo {author}
  {\bibfnamefont {Germ\'an~Fern\'andez}\ \bibnamefont {Barranco}}, \bibinfo
  {author} {\bibfnamefont {Reinhold}\ \bibnamefont {Flatscher}}, \bibinfo
  {author} {\bibfnamefont {Frank}\ \bibnamefont {Flechtner}}, \bibinfo {author}
  {\bibfnamefont {William~M.}\ \bibnamefont {Folkner}}, \bibinfo {author}
  {\bibfnamefont {Samuel}\ \bibnamefont {Francis}}, \bibinfo {author}
  {\bibfnamefont {Martin~S.}\ \bibnamefont {Gilbert}}, \bibinfo {author}
  {\bibfnamefont {Frank}\ \bibnamefont {Gilles}}, \bibinfo {author}
  {\bibfnamefont {Martin}\ \bibnamefont {Gohlke}}, \bibinfo {author}
  {\bibfnamefont {Nicolas}\ \bibnamefont {Grossard}}, \bibinfo {author}
  {\bibfnamefont {Burghardt}\ \bibnamefont {Guenther}}, \bibinfo {author}
  {\bibfnamefont {Philipp}\ \bibnamefont {Hager}}, \bibinfo {author}
  {\bibfnamefont {Jerome}\ \bibnamefont {Hauden}}, \bibinfo {author}
  {\bibfnamefont {Frank}\ \bibnamefont {Heine}}, \bibinfo {author}
  {\bibfnamefont {Gerhard}\ \bibnamefont {Heinzel}}, \bibinfo {author}
  {\bibfnamefont {Mark}\ \bibnamefont {Herding}}, \bibinfo {author}
  {\bibfnamefont {Martin}\ \bibnamefont {Hinz}}, \bibinfo {author}
  {\bibfnamefont {James}\ \bibnamefont {Howell}}, \bibinfo {author}
  {\bibfnamefont {Mark}\ \bibnamefont {Katsumura}}, \bibinfo {author}
  {\bibfnamefont {Marina}\ \bibnamefont {Kaufer}}, \bibinfo {author}
  {\bibfnamefont {William}\ \bibnamefont {Klipstein}}, \bibinfo {author}
  {\bibfnamefont {Alexander}\ \bibnamefont {Koch}}, \bibinfo {author}
  {\bibfnamefont {Micah}\ \bibnamefont {Kruger}}, \bibinfo {author}
  {\bibfnamefont {Kameron}\ \bibnamefont {Larsen}}, \bibinfo {author}
  {\bibfnamefont {Anton}\ \bibnamefont {Lebeda}}, \bibinfo {author}
  {\bibfnamefont {Arnold}\ \bibnamefont {Lebeda}}, \bibinfo {author}
  {\bibfnamefont {Thomas}\ \bibnamefont {Leikert}}, \bibinfo {author}
  {\bibfnamefont {Carl~Christian}\ \bibnamefont {Liebe}}, \bibinfo {author}
  {\bibfnamefont {Jehhal}\ \bibnamefont {Liu}}, \bibinfo {author}
  {\bibfnamefont {Lynette}\ \bibnamefont {Lobmeyer}}, \bibinfo {author}
  {\bibfnamefont {Christoph}\ \bibnamefont {Mahrdt}}, \bibinfo {author}
  {\bibfnamefont {Thomas}\ \bibnamefont {Mangoldt}}, \bibinfo {author}
  {\bibfnamefont {Kirk}\ \bibnamefont {McKenzie}}, \bibinfo {author}
  {\bibfnamefont {Malte}\ \bibnamefont {Misfeldt}}, \bibinfo {author}
  {\bibfnamefont {Phillip~R.}\ \bibnamefont {Morton}}, \bibinfo {author}
  {\bibfnamefont {Vitali}\ \bibnamefont {M\"uller}}, \bibinfo {author}
  {\bibfnamefont {Alexander~T.}\ \bibnamefont {Murray}}, \bibinfo {author}
  {\bibfnamefont {Don~J.}\ \bibnamefont {Nguyen}}, \bibinfo {author}
  {\bibfnamefont {Kolja}\ \bibnamefont {Nicklaus}}, \bibinfo {author}
  {\bibfnamefont {Robert}\ \bibnamefont {Pierce}}, \bibinfo {author}
  {\bibfnamefont {Joshua~A.}\ \bibnamefont {Ravich}}, \bibinfo {author}
  {\bibfnamefont {Gretchen}\ \bibnamefont {Reavis}}, \bibinfo {author}
  {\bibfnamefont {Jens}\ \bibnamefont {Reiche}}, \bibinfo {author}
  {\bibfnamefont {Josep}\ \bibnamefont {Sanjuan}}, \bibinfo {author}
  {\bibfnamefont {Daniel}\ \bibnamefont {Sch\"utze}}, \bibinfo {author}
  {\bibfnamefont {Christoph}\ \bibnamefont {Seiter}}, \bibinfo {author}
  {\bibfnamefont {Daniel}\ \bibnamefont {Shaddock}}, \bibinfo {author}
  {\bibfnamefont {Benjamin}\ \bibnamefont {Sheard}}, \bibinfo {author}
  {\bibfnamefont {Michael}\ \bibnamefont {Sileo}}, \bibinfo {author}
  {\bibfnamefont {Robert}\ \bibnamefont {Spero}}, \bibinfo {author}
  {\bibfnamefont {Gary}\ \bibnamefont {Spiers}}, \bibinfo {author}
  {\bibfnamefont {Gunnar}\ \bibnamefont {Stede}}, \bibinfo {author}
  {\bibfnamefont {Michelle}\ \bibnamefont {Stephens}}, \bibinfo {author}
  {\bibfnamefont {Andrew}\ \bibnamefont {Sutton}}, \bibinfo {author}
  {\bibfnamefont {Joseph}\ \bibnamefont {Trinh}}, \bibinfo {author}
  {\bibfnamefont {Kai}\ \bibnamefont {Voss}}, \bibinfo {author} {\bibfnamefont
  {Duo}\ \bibnamefont {Wang}}, \bibinfo {author} {\bibfnamefont {Rabi~T.}\
  \bibnamefont {Wang}}, \bibinfo {author} {\bibfnamefont {Brent}\ \bibnamefont
  {Ware}}, \bibinfo {author} {\bibfnamefont {Henry}\ \bibnamefont {Wegener}},
  \bibinfo {author} {\bibfnamefont {Steve}\ \bibnamefont {Windisch}}, \bibinfo
  {author} {\bibfnamefont {Christopher}\ \bibnamefont {Woodruff}}, \bibinfo
  {author} {\bibfnamefont {Bernd}\ \bibnamefont {Zender}}, \ and\ \bibinfo
  {author} {\bibfnamefont {Marcus}\ \bibnamefont {Zimmermann}},\ }\bibfield
  {title} {\enquote {\bibinfo {title} {In-orbit performance of the grace
  follow-on laser ranging interferometer},}\ }\href {\doibase
  10.1103/PhysRevLett.123.031101} {\bibfield  {journal} {\bibinfo  {journal}
  {Phys. Rev. Lett.}\ }\textbf {\bibinfo {volume} {123}},\ \bibinfo {pages}
  {031101} (\bibinfo {year} {2019})}\BibitemShut {NoStop}%
\bibitem [{\citenamefont {Cruise}\ \emph {et~al.}(2005)\citenamefont {Cruise},
  \citenamefont {Hoyland},\ and\ \citenamefont {Aston}}]{PMLTP2}%
  \BibitemOpen
  \bibfield  {author} {\bibinfo {author} {\bibfnamefont {A.~M.}\ \bibnamefont
  {Cruise}}, \bibinfo {author} {\bibfnamefont {D.}~\bibnamefont {Hoyland}}, \
  and\ \bibinfo {author} {\bibfnamefont {S.~M.}\ \bibnamefont {Aston}},\
  }\bibfield  {title} {\enquote {\bibinfo {title} {Implementation of the
  phasemeter for lisa ltp},}\ }\href {\doibase 10.1088/0264-9381/22/10/005}
  {\bibfield  {journal} {\bibinfo  {journal} {Classical and Quantum Gravity}\
  }\textbf {\bibinfo {volume} {22}},\ \bibinfo {pages} {S165--S169} (\bibinfo
  {year} {2005})}\BibitemShut {NoStop}%
\bibitem [{\citenamefont {Heinzel}\ \emph {et~al.}(2004)\citenamefont
  {Heinzel}, \citenamefont {Wand}, \citenamefont {Garcia}, \citenamefont
  {Jennrich}, \citenamefont {Braxmaier}, \citenamefont {Robertson},
  \citenamefont {Middleton}, \citenamefont {Hoyland}, \citenamefont {Rudiger},
  \citenamefont {Schilling}, \citenamefont {Johann},\ and\ \citenamefont
  {Danzmann}}]{PMLTP1}%
  \BibitemOpen
  \bibfield  {author} {\bibinfo {author} {\bibfnamefont {G.}~\bibnamefont
  {Heinzel}}, \bibinfo {author} {\bibfnamefont {V.}~\bibnamefont {Wand}},
  \bibinfo {author} {\bibfnamefont {A.}~\bibnamefont {Garcia}}, \bibinfo
  {author} {\bibfnamefont {O.~P.}\ \bibnamefont {Jennrich}}, \bibinfo {author}
  {\bibfnamefont {C.}~\bibnamefont {Braxmaier}}, \bibinfo {author}
  {\bibfnamefont {D.}~\bibnamefont {Robertson}}, \bibinfo {author}
  {\bibfnamefont {K.}~\bibnamefont {Middleton}}, \bibinfo {author}
  {\bibfnamefont {D.}~\bibnamefont {Hoyland}}, \bibinfo {author} {\bibfnamefont
  {A.}~\bibnamefont {Rudiger}}, \bibinfo {author} {\bibfnamefont
  {R.}~\bibnamefont {Schilling}}, \bibinfo {author} {\bibfnamefont
  {U.}~\bibnamefont {Johann}}, \ and\ \bibinfo {author} {\bibfnamefont
  {K.}~\bibnamefont {Danzmann}},\ }\bibfield  {title} {\enquote {\bibinfo
  {title} {The ltp interferometer and phasemeter},}\ }\href {\doibase
  10.1088/0264-9381/21/5/029} {\bibfield  {journal} {\bibinfo  {journal}
  {Classical and Quantum Gravity}\ }\textbf {\bibinfo {volume} {21}},\ \bibinfo
  {pages} {S581--S587} (\bibinfo {year} {2004})}\BibitemShut {NoStop}%
\bibitem [{\citenamefont {Liang}\ \emph {et~al.}(2012)\citenamefont {Liang},
  \citenamefont {Duan}, \citenamefont {Yeh},\ and\ \citenamefont
  {Luo}}]{PMCROSS}%
  \BibitemOpen
  \bibfield  {author} {\bibinfo {author} {\bibfnamefont {Yu-Rong}\ \bibnamefont
  {Liang}}, \bibinfo {author} {\bibfnamefont {Hui-Zong}\ \bibnamefont {Duan}},
  \bibinfo {author} {\bibfnamefont {Hsien-Chi}\ \bibnamefont {Yeh}}, \ and\
  \bibinfo {author} {\bibfnamefont {Jun}\ \bibnamefont {Luo}},\ }\bibfield
  {title} {\enquote {\bibinfo {title} {Fundamental limits on the digital phase
  measurement method based on cross-correlation analysis},}\ }\href {\doibase
  10.1063/1.4751867} {\bibfield  {journal} {\bibinfo  {journal} {Review of
  Scientific Instruments}\ }\textbf {\bibinfo {volume} {83}} (\bibinfo {year}
  {2012}),\ 10.1063/1.4751867}\BibitemShut {NoStop}%
\bibitem [{\citenamefont {Pollack}\ and\ \citenamefont
  {Stebbins}(2006)}]{PMZERO}%
  \BibitemOpen
  \bibfield  {author} {\bibinfo {author} {\bibfnamefont {S.~E.}\ \bibnamefont
  {Pollack}}\ and\ \bibinfo {author} {\bibfnamefont {R.~T.}\ \bibnamefont
  {Stebbins}},\ }\bibfield  {title} {\enquote {\bibinfo {title} {Demonstration
  of the zero-crossing phasemeter with a lisa test-bed interferometer},}\
  }\href {\doibase 10.1088/0264-9381/23/12/014} {\bibfield  {journal} {\bibinfo
   {journal} {Classical and Quantum Gravity}\ }\textbf {\bibinfo {volume}
  {23}},\ \bibinfo {pages} {4189--4200} (\bibinfo {year} {2006})}\BibitemShut
  {NoStop}%
\bibitem [{\citenamefont {Wand}\ \emph {et~al.}(2006)\citenamefont {Wand},
  \citenamefont {Guzman}, \citenamefont {Heinzel},\ and\ \citenamefont
  {Danzmann}}]{PMLISA1}%
  \BibitemOpen
  \bibfield  {author} {\bibinfo {author} {\bibfnamefont {Vinzenz}\ \bibnamefont
  {Wand}}, \bibinfo {author} {\bibfnamefont {Felipe}\ \bibnamefont {Guzman}},
  \bibinfo {author} {\bibfnamefont {Gerhard}\ \bibnamefont {Heinzel}}, \ and\
  \bibinfo {author} {\bibfnamefont {Karsten}\ \bibnamefont {Danzmann}},\
  }\enquote {\bibinfo {title} {Lisa phasemeter development},}\ in\ \href
  {https://doi.org/10.1063/1.2405118} {\emph {\bibinfo {booktitle} {Laser
  Interferometer Space Antenna}}},\ \bibinfo {series} {AIP Conference
  Proceedings}, Vol.\ \bibinfo {volume} {873},\ \bibinfo {editor} {edited by\
  \bibinfo {editor} {\bibfnamefont {S.~M.}\ \bibnamefont {Merkowitz}}\ and\
  \bibinfo {editor} {\bibfnamefont {J.~C.}\ \bibnamefont {Livas}}}\ (\bibinfo
  {year} {2006})\ pp.\ \bibinfo {pages} {689--+}\BibitemShut {NoStop}%
\bibitem [{\citenamefont {Shaddock}\ \emph {et~al.}(2006)\citenamefont
  {Shaddock}, \citenamefont {Ware}, \citenamefont {Halverson}, \citenamefont
  {Spero},\ and\ \citenamefont {Klipstein}}]{PMLISA2}%
  \BibitemOpen
  \bibfield  {author} {\bibinfo {author} {\bibfnamefont {D.}~\bibnamefont
  {Shaddock}}, \bibinfo {author} {\bibfnamefont {B.}~\bibnamefont {Ware}},
  \bibinfo {author} {\bibfnamefont {P.~G.}\ \bibnamefont {Halverson}}, \bibinfo
  {author} {\bibfnamefont {R.~E.}\ \bibnamefont {Spero}}, \ and\ \bibinfo
  {author} {\bibfnamefont {B.}~\bibnamefont {Klipstein}},\ }\enquote {\bibinfo
  {title} {Overview of the lisa phasemeter},}\ in\ \href {\doibase
  10.1063/1.2405113} {\emph {\bibinfo {booktitle} {Laser Interferometer Space
  Antenna}}},\ \bibinfo {series} {AIP Conference Proceedings}, Vol.\ \bibinfo
  {volume} {873},\ \bibinfo {editor} {edited by\ \bibinfo {editor}
  {\bibfnamefont {S.~M.}\ \bibnamefont {Merkowitz}}\ and\ \bibinfo {editor}
  {\bibfnamefont {J.~C.}\ \bibnamefont {Livas}}}\ (\bibinfo {year} {2006})\
  pp.\ \bibinfo {pages} {654--+}\BibitemShut {NoStop}%
\bibitem [{\citenamefont {Gray}\ \emph {et~al.}(2012)\citenamefont {Gray},
  \citenamefont {McRae}, \citenamefont {Hsu}, \citenamefont {Herrmann},\ and\
  \citenamefont {Shaddock}}]{PMLISA3}%
  \BibitemOpen
  \bibfield  {author} {\bibinfo {author} {\bibfnamefont {Malcolm~B.}\
  \bibnamefont {Gray}}, \bibinfo {author} {\bibfnamefont {Terry}\ \bibnamefont
  {McRae}}, \bibinfo {author} {\bibfnamefont {Magnus T.~L.}\ \bibnamefont
  {Hsu}}, \bibinfo {author} {\bibfnamefont {Jan}\ \bibnamefont {Herrmann}}, \
  and\ \bibinfo {author} {\bibfnamefont {Daniel~A.}\ \bibnamefont {Shaddock}},\
  }\enquote {\bibinfo {title} {A digital phasemeter for precision length
  measurements},}\ in\ \href {\doibase 10.1364/CLEO_AT.2012.JW1C.4} {\emph
  {\bibinfo {booktitle} {2012 Conference on Lasers and Electro-Optics}}},\
  \bibinfo {series and number} {Conference on Lasers and Electro-Optics}\
  (\bibinfo {year} {2012})\BibitemShut {NoStop}%
\bibitem [{\citenamefont {Gerberding}\ \emph {et~al.}(2013)\citenamefont
  {Gerberding}, \citenamefont {Sheard}, \citenamefont {Bykov}, \citenamefont
  {Kullmann}, \citenamefont {Delgado}, \citenamefont {Danzmann},\ and\
  \citenamefont {Heinzel}}]{PMLISA4}%
  \BibitemOpen
  \bibfield  {author} {\bibinfo {author} {\bibfnamefont {Oliver}\ \bibnamefont
  {Gerberding}}, \bibinfo {author} {\bibfnamefont {Benjamin}\ \bibnamefont
  {Sheard}}, \bibinfo {author} {\bibfnamefont {Iouri}\ \bibnamefont {Bykov}},
  \bibinfo {author} {\bibfnamefont {Joachim}\ \bibnamefont {Kullmann}},
  \bibinfo {author} {\bibfnamefont {Juan Jose~Esteban}\ \bibnamefont
  {Delgado}}, \bibinfo {author} {\bibfnamefont {Karsten}\ \bibnamefont
  {Danzmann}}, \ and\ \bibinfo {author} {\bibfnamefont {Gerhard}\ \bibnamefont
  {Heinzel}},\ }\bibfield  {title} {\enquote {\bibinfo {title} {Phasemeter core
  for intersatellite laser heterodyne interferometry: modelling, simulations
  and experiments},}\ }\href {\doibase 10.1088/0264-9381/30/23/235029}
  {\bibfield  {journal} {\bibinfo  {journal} {Classical and Quantum Gravity}\
  }\textbf {\bibinfo {volume} {30}} (\bibinfo {year} {2013}),\
  10.1088/0264-9381/30/23/235029}\BibitemShut {NoStop}%
\bibitem [{\citenamefont {Francis}\ \emph {et~al.}(2014)\citenamefont
  {Francis}, \citenamefont {Lam}, \citenamefont {McKenzie}, \citenamefont
  {Sutton}, \citenamefont {Ward}, \citenamefont {McClelland},\ and\
  \citenamefont {Shaddock}}]{PMLISA5}%
  \BibitemOpen
  \bibfield  {author} {\bibinfo {author} {\bibfnamefont {Samuel~P.}\
  \bibnamefont {Francis}}, \bibinfo {author} {\bibfnamefont {Timothy T.~Y.}\
  \bibnamefont {Lam}}, \bibinfo {author} {\bibfnamefont {Kirk}\ \bibnamefont
  {McKenzie}}, \bibinfo {author} {\bibfnamefont {Andrew~J.}\ \bibnamefont
  {Sutton}}, \bibinfo {author} {\bibfnamefont {Robert~L.}\ \bibnamefont
  {Ward}}, \bibinfo {author} {\bibfnamefont {David~E.}\ \bibnamefont
  {McClelland}}, \ and\ \bibinfo {author} {\bibfnamefont {Daniel~A.}\
  \bibnamefont {Shaddock}},\ }\bibfield  {title} {\enquote {\bibinfo {title}
  {Weak-light phase tracking with a low cycle slip rate},}\ }\href {\doibase
  10.1364/ol.39.005251} {\bibfield  {journal} {\bibinfo  {journal} {Optics
  Letters}\ }\textbf {\bibinfo {volume} {39}},\ \bibinfo {pages} {5251--5254}
  (\bibinfo {year} {2014})}\BibitemShut {NoStop}%
\bibitem [{\citenamefont {Bachman}\ \emph {et~al.}(2017)\citenamefont
  {Bachman}, \citenamefont {de~Vine}, \citenamefont {Dickson}, \citenamefont
  {Dubovitsky}, \citenamefont {Liu}, \citenamefont {Klipstein}, \citenamefont
  {McKenzie}, \citenamefont {Spero}, \citenamefont {Sutton}, \citenamefont
  {Ware},\ and\ \citenamefont {Woodruff}}]{PMLRI}%
  \BibitemOpen
  \bibfield  {author} {\bibinfo {author} {\bibfnamefont {B.}~\bibnamefont
  {Bachman}}, \bibinfo {author} {\bibfnamefont {G.}~\bibnamefont {de~Vine}},
  \bibinfo {author} {\bibfnamefont {J.}~\bibnamefont {Dickson}}, \bibinfo
  {author} {\bibfnamefont {S.}~\bibnamefont {Dubovitsky}}, \bibinfo {author}
  {\bibfnamefont {J.}~\bibnamefont {Liu}}, \bibinfo {author} {\bibfnamefont
  {W.}~\bibnamefont {Klipstein}}, \bibinfo {author} {\bibfnamefont
  {K.}~\bibnamefont {McKenzie}}, \bibinfo {author} {\bibfnamefont
  {R.}~\bibnamefont {Spero}}, \bibinfo {author} {\bibfnamefont
  {A.}~\bibnamefont {Sutton}}, \bibinfo {author} {\bibfnamefont
  {B.}~\bibnamefont {Ware}}, \ and\ \bibinfo {author} {\bibfnamefont
  {C.}~\bibnamefont {Woodruff}},\ }\bibfield  {title} {\enquote {\bibinfo
  {title} {Flight phasemeter on the laser ranging interferometer on the grace
  follow-on mission},}\ }in\ \href {\doibase 10.1088/1742-6596/840/1/012011}
  {\emph {\bibinfo {booktitle} {11th International LISA Symposium}}},\ \bibinfo
  {series} {Journal of Physics Conference Series}, Vol.\ \bibinfo {volume}
  {840}\ (\bibinfo {year} {2017})\BibitemShut {NoStop}%
\bibitem [{\citenamefont {Danzmann}\ \emph {et~al.}(2015)\citenamefont
  {Danzmann}, \citenamefont {{Lisa Pathfinder Team}},\ and\ \citenamefont
  {{eLisa Consortium}}}]{LISA}%
  \BibitemOpen
  \bibfield  {author} {\bibinfo {author} {\bibfnamefont {Karsten}\ \bibnamefont
  {Danzmann}}, \bibinfo {author} {\bibnamefont {{Lisa Pathfinder Team}}}, \
  and\ \bibinfo {author} {\bibnamefont {{eLisa Consortium}}},\ }\bibfield
  {title} {\enquote {\bibinfo {title} {Lisa and its pathfinder},}\ }\href
  {\doibase 10.1038/nphys3420} {\bibfield  {journal} {\bibinfo  {journal}
  {Nature Physics}\ }\textbf {\bibinfo {volume} {11}},\ \bibinfo {pages}
  {613--615} (\bibinfo {year} {2015})}\BibitemShut {NoStop}%
\bibitem [{L3P()}]{L3Proposal}%
  \BibitemOpen
  \href {https://www.elisascience.org/files/publications/LISA_L3_20170120.pdf}
  {\enquote {\bibinfo {title} {{LISA} mission {L}3 proposal},}\ }\BibitemShut
  {NoStop}%
\bibitem [{\citenamefont {Sheard}\ \emph {et~al.}(2012)\citenamefont {Sheard},
  \citenamefont {Heinzel}, \citenamefont {Danzmann}, \citenamefont {Shaddock},
  \citenamefont {Klipstein},\ and\ \citenamefont {Folkner}}]{LRI}%
  \BibitemOpen
  \bibfield  {author} {\bibinfo {author} {\bibfnamefont {B.~S.}\ \bibnamefont
  {Sheard}}, \bibinfo {author} {\bibfnamefont {G.}~\bibnamefont {Heinzel}},
  \bibinfo {author} {\bibfnamefont {K.}~\bibnamefont {Danzmann}}, \bibinfo
  {author} {\bibfnamefont {D.~A.}\ \bibnamefont {Shaddock}}, \bibinfo {author}
  {\bibfnamefont {W.~M.}\ \bibnamefont {Klipstein}}, \ and\ \bibinfo {author}
  {\bibfnamefont {W.~M.}\ \bibnamefont {Folkner}},\ }\bibfield  {title}
  {\enquote {\bibinfo {title} {Intersatellite laser ranging instrument for the
  grace follow-on mission},}\ }\href {\doibase 10.1007/s00190-012-0566-3}
  {\bibfield  {journal} {\bibinfo  {journal} {Journal of Geodesy}\ }\textbf
  {\bibinfo {volume} {86}},\ \bibinfo {pages} {1083--1095} (\bibinfo {year}
  {2012})}\BibitemShut {NoStop}%
\bibitem [{LRI()}]{LRIRES}%
  \BibitemOpen
  \href@noop {} {\enquote {\bibinfo {title} {{Joint press release by AEI and
  NASA/JPL, July 02, 2018}; paper in preparation},}\ }\BibitemShut {NoStop}%
\bibitem [{\citenamefont {Gerberding}(2014)}]{OLI}%
  \BibitemOpen
  \bibfield  {author} {\bibinfo {author} {\bibfnamefont {Oliver}\ \bibnamefont
  {Gerberding}},\ }\emph {\bibinfo {title} {Phase readout for satellite
  interferometry}},\ \href
  {http://edok01.tib.uni-hannover.de/edoks/e01dh14/783659903.pdf} {Ph.D.
  thesis},\ \bibinfo  {school} {Leibniz Universit\"at Hannover} (\bibinfo
  {year} {2014})\BibitemShut {NoStop}%
\bibitem [{\citenamefont {Armano}\ \emph {et~al.}(2016)\citenamefont {Armano},
  \citenamefont {Audley}, \citenamefont {Auger}, \citenamefont {Baird},
  \citenamefont {Bassan}, \citenamefont {Binetruy}, \citenamefont {Born},
  \citenamefont {Bortoluzzi}, \citenamefont {Brandt}, \citenamefont {Caleno},
  \citenamefont {Carbone}, \citenamefont {Cavalleri}, \citenamefont {Cesarini},
  \citenamefont {Ciani}, \citenamefont {Congedo}, \citenamefont {Cruise},
  \citenamefont {Danzmann}, \citenamefont {de~Deus~Silva}, \citenamefont
  {De~Rosa}, \citenamefont {Diaz-Aguilo}, \citenamefont {Di~Fiore},
  \citenamefont {Diepholz}, \citenamefont {Dixon}, \citenamefont {Dolesi},
  \citenamefont {Dunbar}, \citenamefont {Ferraioli}, \citenamefont {Ferroni},
  \citenamefont {Fichter}, \citenamefont {Fitzsimons}, \citenamefont
  {Flatscher}, \citenamefont {Freschi}, \citenamefont {Marin}, \citenamefont
  {Marirrodriga}, \citenamefont {Gerndt}, \citenamefont {Gesa}, \citenamefont
  {Gibert}, \citenamefont {Giardini}, \citenamefont {Giusteri}, \citenamefont
  {Guzman}, \citenamefont {Grado}, \citenamefont {Grimani}, \citenamefont
  {Grynagier}, \citenamefont {Grzymisch}, \citenamefont {Harrison},
  \citenamefont {Heinzel}, \citenamefont {Hewitson}, \citenamefont
  {Hollington}, \citenamefont {Hoyland}, \citenamefont {Hueller}, \citenamefont
  {Inchauspe}, \citenamefont {Jennrich}, \citenamefont {Jetzer}, \citenamefont
  {Johann}, \citenamefont {Johlander}, \citenamefont {Karnesis}, \citenamefont
  {Kaune}, \citenamefont {Korsakova}, \citenamefont {Killow}, \citenamefont
  {Lobo}, \citenamefont {Lloro}, \citenamefont {Liu}, \citenamefont
  {Lopez-Zaragoza}, \citenamefont {Maarschalkerweerd}, \citenamefont {Mance},
  \citenamefont {Martin}, \citenamefont {Martin-Polo}, \citenamefont {Martino},
  \citenamefont {Martin-Porqueras}, \citenamefont {Madden}, \citenamefont
  {Mateos}, \citenamefont {McNamara}, \citenamefont {Mendes}, \citenamefont
  {Mendes}, \citenamefont {Monsky}, \citenamefont {Nicolodi}, \citenamefont
  {Nofrarias}, \citenamefont {Paczkowski}, \citenamefont {Perreur-Lloyd},
  \citenamefont {Petiteau}, \citenamefont {Pivato}, \citenamefont {Plagnol},
  \citenamefont {Prat}, \citenamefont {Ragnit}, \citenamefont {Rais},
  \citenamefont {Ramos-Castro}, \citenamefont {Reiche}, \citenamefont
  {Robertson}, \citenamefont {Rozemeijer}, \citenamefont {Rivas}, \citenamefont
  {Russano}, \citenamefont {Sanjuan}, \citenamefont {Sarra}, \citenamefont
  {Schleicher}, \citenamefont {Shaul}, \citenamefont {Slutsky}, \citenamefont
  {Sopuerta}, \citenamefont {Stanga}, \citenamefont {Steier}, \citenamefont
  {Sumner}, \citenamefont {Texier} \emph {et~al.}}]{LTP1}%
  \BibitemOpen
  \bibfield  {author} {\bibinfo {author} {\bibfnamefont {M.}~\bibnamefont
  {Armano}}, \bibinfo {author} {\bibfnamefont {H.}~\bibnamefont {Audley}},
  \bibinfo {author} {\bibfnamefont {G.}~\bibnamefont {Auger}}, \bibinfo
  {author} {\bibfnamefont {J.~T.}\ \bibnamefont {Baird}}, \bibinfo {author}
  {\bibfnamefont {M.}~\bibnamefont {Bassan}}, \bibinfo {author} {\bibfnamefont
  {P.}~\bibnamefont {Binetruy}}, \bibinfo {author} {\bibfnamefont
  {M.}~\bibnamefont {Born}}, \bibinfo {author} {\bibfnamefont {D.}~\bibnamefont
  {Bortoluzzi}}, \bibinfo {author} {\bibfnamefont {N.}~\bibnamefont {Brandt}},
  \bibinfo {author} {\bibfnamefont {M.}~\bibnamefont {Caleno}}, \bibinfo
  {author} {\bibfnamefont {L.}~\bibnamefont {Carbone}}, \bibinfo {author}
  {\bibfnamefont {A.}~\bibnamefont {Cavalleri}}, \bibinfo {author}
  {\bibfnamefont {A.}~\bibnamefont {Cesarini}}, \bibinfo {author}
  {\bibfnamefont {G.}~\bibnamefont {Ciani}}, \bibinfo {author} {\bibfnamefont
  {G.}~\bibnamefont {Congedo}}, \bibinfo {author} {\bibfnamefont {A.~M.}\
  \bibnamefont {Cruise}}, \bibinfo {author} {\bibfnamefont {K.}~\bibnamefont
  {Danzmann}}, \bibinfo {author} {\bibfnamefont {M.}~\bibnamefont
  {de~Deus~Silva}}, \bibinfo {author} {\bibfnamefont {R.}~\bibnamefont
  {De~Rosa}}, \bibinfo {author} {\bibfnamefont {M.}~\bibnamefont
  {Diaz-Aguilo}}, \bibinfo {author} {\bibfnamefont {L.}~\bibnamefont
  {Di~Fiore}}, \bibinfo {author} {\bibfnamefont {I.}~\bibnamefont {Diepholz}},
  \bibinfo {author} {\bibfnamefont {G.}~\bibnamefont {Dixon}}, \bibinfo
  {author} {\bibfnamefont {R.}~\bibnamefont {Dolesi}}, \bibinfo {author}
  {\bibfnamefont {N.}~\bibnamefont {Dunbar}}, \bibinfo {author} {\bibfnamefont
  {L.}~\bibnamefont {Ferraioli}}, \bibinfo {author} {\bibfnamefont
  {V.}~\bibnamefont {Ferroni}}, \bibinfo {author} {\bibfnamefont
  {W.}~\bibnamefont {Fichter}}, \bibinfo {author} {\bibfnamefont {E.~D.}\
  \bibnamefont {Fitzsimons}}, \bibinfo {author} {\bibfnamefont
  {R.}~\bibnamefont {Flatscher}}, \bibinfo {author} {\bibfnamefont
  {M.}~\bibnamefont {Freschi}}, \bibinfo {author} {\bibfnamefont
  {A.~F.~Garcia}\ \bibnamefont {Marin}}, \bibinfo {author} {\bibfnamefont
  {C.~Garcia}\ \bibnamefont {Marirrodriga}}, \bibinfo {author} {\bibfnamefont
  {R.}~\bibnamefont {Gerndt}}, \bibinfo {author} {\bibfnamefont
  {L.}~\bibnamefont {Gesa}}, \bibinfo {author} {\bibfnamefont {F.}~\bibnamefont
  {Gibert}}, \bibinfo {author} {\bibfnamefont {D.}~\bibnamefont {Giardini}},
  \bibinfo {author} {\bibfnamefont {R.}~\bibnamefont {Giusteri}}, \bibinfo
  {author} {\bibfnamefont {F.}~\bibnamefont {Guzman}}, \bibinfo {author}
  {\bibfnamefont {A.}~\bibnamefont {Grado}}, \bibinfo {author} {\bibfnamefont
  {C.}~\bibnamefont {Grimani}}, \bibinfo {author} {\bibfnamefont
  {A.}~\bibnamefont {Grynagier}}, \bibinfo {author} {\bibfnamefont
  {J.}~\bibnamefont {Grzymisch}}, \bibinfo {author} {\bibfnamefont
  {I.}~\bibnamefont {Harrison}}, \bibinfo {author} {\bibfnamefont
  {G.}~\bibnamefont {Heinzel}}, \bibinfo {author} {\bibfnamefont
  {M.}~\bibnamefont {Hewitson}}, \bibinfo {author} {\bibfnamefont
  {D.}~\bibnamefont {Hollington}}, \bibinfo {author} {\bibfnamefont
  {D.}~\bibnamefont {Hoyland}}, \bibinfo {author} {\bibfnamefont
  {M.}~\bibnamefont {Hueller}}, \bibinfo {author} {\bibfnamefont
  {H.}~\bibnamefont {Inchauspe}}, \bibinfo {author} {\bibfnamefont
  {O.}~\bibnamefont {Jennrich}}, \bibinfo {author} {\bibfnamefont
  {P.}~\bibnamefont {Jetzer}}, \bibinfo {author} {\bibfnamefont
  {U.}~\bibnamefont {Johann}}, \bibinfo {author} {\bibfnamefont
  {B.}~\bibnamefont {Johlander}}, \bibinfo {author} {\bibfnamefont
  {N.}~\bibnamefont {Karnesis}}, \bibinfo {author} {\bibfnamefont
  {B.}~\bibnamefont {Kaune}}, \bibinfo {author} {\bibfnamefont
  {N.}~\bibnamefont {Korsakova}}, \bibinfo {author} {\bibfnamefont {C.~J.}\
  \bibnamefont {Killow}}, \bibinfo {author} {\bibfnamefont {J.~A.}\
  \bibnamefont {Lobo}}, \bibinfo {author} {\bibfnamefont {I.}~\bibnamefont
  {Lloro}}, \bibinfo {author} {\bibfnamefont {L.}~\bibnamefont {Liu}}, \bibinfo
  {author} {\bibfnamefont {J.~P.}\ \bibnamefont {Lopez-Zaragoza}}, \bibinfo
  {author} {\bibfnamefont {R.}~\bibnamefont {Maarschalkerweerd}}, \bibinfo
  {author} {\bibfnamefont {D.}~\bibnamefont {Mance}}, \bibinfo {author}
  {\bibfnamefont {V.}~\bibnamefont {Martin}}, \bibinfo {author} {\bibfnamefont
  {L.}~\bibnamefont {Martin-Polo}}, \bibinfo {author} {\bibfnamefont
  {J.}~\bibnamefont {Martino}}, \bibinfo {author} {\bibfnamefont
  {F.}~\bibnamefont {Martin-Porqueras}}, \bibinfo {author} {\bibfnamefont
  {S.}~\bibnamefont {Madden}}, \bibinfo {author} {\bibfnamefont
  {I.}~\bibnamefont {Mateos}}, \bibinfo {author} {\bibfnamefont {P.~W.}\
  \bibnamefont {McNamara}}, \bibinfo {author} {\bibfnamefont {J.}~\bibnamefont
  {Mendes}}, \bibinfo {author} {\bibfnamefont {L.}~\bibnamefont {Mendes}},
  \bibinfo {author} {\bibfnamefont {A.}~\bibnamefont {Monsky}}, \bibinfo
  {author} {\bibfnamefont {D.}~\bibnamefont {Nicolodi}}, \bibinfo {author}
  {\bibfnamefont {M.}~\bibnamefont {Nofrarias}}, \bibinfo {author}
  {\bibfnamefont {S.}~\bibnamefont {Paczkowski}}, \bibinfo {author}
  {\bibfnamefont {M.}~\bibnamefont {Perreur-Lloyd}}, \bibinfo {author}
  {\bibfnamefont {A.}~\bibnamefont {Petiteau}}, \bibinfo {author}
  {\bibfnamefont {P.}~\bibnamefont {Pivato}}, \bibinfo {author} {\bibfnamefont
  {E.}~\bibnamefont {Plagnol}}, \bibinfo {author} {\bibfnamefont
  {P.}~\bibnamefont {Prat}}, \bibinfo {author} {\bibfnamefont {U.}~\bibnamefont
  {Ragnit}}, \bibinfo {author} {\bibfnamefont {B.}~\bibnamefont {Rais}},
  \bibinfo {author} {\bibfnamefont {J.}~\bibnamefont {Ramos-Castro}}, \bibinfo
  {author} {\bibfnamefont {J.}~\bibnamefont {Reiche}}, \bibinfo {author}
  {\bibfnamefont {D.~I.}\ \bibnamefont {Robertson}}, \bibinfo {author}
  {\bibfnamefont {H.}~\bibnamefont {Rozemeijer}}, \bibinfo {author}
  {\bibfnamefont {F.}~\bibnamefont {Rivas}}, \bibinfo {author} {\bibfnamefont
  {G.}~\bibnamefont {Russano}}, \bibinfo {author} {\bibfnamefont
  {J.}~\bibnamefont {Sanjuan}}, \bibinfo {author} {\bibfnamefont
  {P.}~\bibnamefont {Sarra}}, \bibinfo {author} {\bibfnamefont
  {A.}~\bibnamefont {Schleicher}}, \bibinfo {author} {\bibfnamefont
  {D.}~\bibnamefont {Shaul}}, \bibinfo {author} {\bibfnamefont
  {J.}~\bibnamefont {Slutsky}}, \bibinfo {author} {\bibfnamefont {C.~F.}\
  \bibnamefont {Sopuerta}}, \bibinfo {author} {\bibfnamefont {R.}~\bibnamefont
  {Stanga}}, \bibinfo {author} {\bibfnamefont {F.}~\bibnamefont {Steier}},
  \bibinfo {author} {\bibfnamefont {T.}~\bibnamefont {Sumner}}, \bibinfo
  {author} {\bibfnamefont {D.}~\bibnamefont {Texier}},  \emph {et~al.},\
  }\bibfield  {title} {\enquote {\bibinfo {title} {Sub-femto-g free fall for
  space-based gravitational wave observatories: Lisa pathfinder results},}\
  }\href {\doibase 10.1103/PhysRevLett.116.231101} {\bibfield  {journal}
  {\bibinfo  {journal} {Physical Review Letters}\ }\textbf {\bibinfo {volume}
  {116}} (\bibinfo {year} {2016}),\ 10.1103/PhysRevLett.116.231101}\BibitemShut
  {NoStop}%
\bibitem [{\citenamefont {Armano}\ \emph {et~al.}(2019)\citenamefont {Armano},
  \citenamefont {Audley}, \citenamefont {Baird}, \citenamefont {Binetruy},
  \citenamefont {Born}, \citenamefont {Bortoluzzi}, \citenamefont {Castelli},
  \citenamefont {Cavalleri}, \citenamefont {Cesarini}, \citenamefont {Cruise},
  \citenamefont {Danzmann}, \citenamefont {de~Deus~Silva}, \citenamefont
  {Diepholz}, \citenamefont {Dixon}, \citenamefont {Dolesi}, \citenamefont
  {Ferraioli}, \citenamefont {Ferroni}, \citenamefont {Fitzsimons},
  \citenamefont {Freschi}, \citenamefont {Gesa}, \citenamefont {Gibert},
  \citenamefont {Giardini}, \citenamefont {Giusteri}, \citenamefont {Grimani},
  \citenamefont {Grzymisch}, \citenamefont {Harrison}, \citenamefont {Heinzel},
  \citenamefont {Hewitson}, \citenamefont {Hollington}, \citenamefont
  {Hoyland}, \citenamefont {Hueller}, \citenamefont {Inchausp\'e},
  \citenamefont {Jennrich}, \citenamefont {Jetzer}, \citenamefont {Karnesis},
  \citenamefont {Kaune}, \citenamefont {Korsakova}, \citenamefont {Killow},
  \citenamefont {Lobo}, \citenamefont {Lloro}, \citenamefont {Liu},
  \citenamefont {L\'opez-Zaragoza}, \citenamefont {Maarschalkerweerd},
  \citenamefont {Mance}, \citenamefont {Meshksar}, \citenamefont {Mart\'{\i}n},
  \citenamefont {Martin-Polo}, \citenamefont {Martino}, \citenamefont
  {Martin-Porqueras}, \citenamefont {Mateos}, \citenamefont {McNamara},
  \citenamefont {Mendes}, \citenamefont {Mendes}, \citenamefont {Nofrarias},
  \citenamefont {Paczkowski}, \citenamefont {Perreur-Lloyd}, \citenamefont
  {Petiteau}, \citenamefont {Pivato}, \citenamefont {Plagnol}, \citenamefont
  {Ramos-Castro}, \citenamefont {Reiche}, \citenamefont {Robertson},
  \citenamefont {Rivas}, \citenamefont {Russano}, \citenamefont {Slutsky},
  \citenamefont {Sopuerta}, \citenamefont {Sumner}, \citenamefont {Texier},
  \citenamefont {Thorpe}, \citenamefont {Vetrugno}, \citenamefont {Vitale},
  \citenamefont {Wanner}, \citenamefont {Ward}, \citenamefont {Wass},
  \citenamefont {Weber}, \citenamefont {Wissel}, \citenamefont {Wittchen},\
  and\ \citenamefont {Zweifel}}]{PhysRevD.99.082001}%
  \BibitemOpen
  \bibfield  {author} {\bibinfo {author} {\bibfnamefont {M.}~\bibnamefont
  {Armano}}, \bibinfo {author} {\bibfnamefont {H.}~\bibnamefont {Audley}},
  \bibinfo {author} {\bibfnamefont {J.}~\bibnamefont {Baird}}, \bibinfo
  {author} {\bibfnamefont {P.}~\bibnamefont {Binetruy}}, \bibinfo {author}
  {\bibfnamefont {M.}~\bibnamefont {Born}}, \bibinfo {author} {\bibfnamefont
  {D.}~\bibnamefont {Bortoluzzi}}, \bibinfo {author} {\bibfnamefont
  {E.}~\bibnamefont {Castelli}}, \bibinfo {author} {\bibfnamefont
  {A.}~\bibnamefont {Cavalleri}}, \bibinfo {author} {\bibfnamefont
  {A.}~\bibnamefont {Cesarini}}, \bibinfo {author} {\bibfnamefont {A.~M.}\
  \bibnamefont {Cruise}}, \bibinfo {author} {\bibfnamefont {K.}~\bibnamefont
  {Danzmann}}, \bibinfo {author} {\bibfnamefont {M.}~\bibnamefont
  {de~Deus~Silva}}, \bibinfo {author} {\bibfnamefont {I.}~\bibnamefont
  {Diepholz}}, \bibinfo {author} {\bibfnamefont {G.}~\bibnamefont {Dixon}},
  \bibinfo {author} {\bibfnamefont {R.}~\bibnamefont {Dolesi}}, \bibinfo
  {author} {\bibfnamefont {L.}~\bibnamefont {Ferraioli}}, \bibinfo {author}
  {\bibfnamefont {V.}~\bibnamefont {Ferroni}}, \bibinfo {author} {\bibfnamefont
  {E.~D.}\ \bibnamefont {Fitzsimons}}, \bibinfo {author} {\bibfnamefont
  {M.}~\bibnamefont {Freschi}}, \bibinfo {author} {\bibfnamefont
  {L.}~\bibnamefont {Gesa}}, \bibinfo {author} {\bibfnamefont {F.}~\bibnamefont
  {Gibert}}, \bibinfo {author} {\bibfnamefont {D.}~\bibnamefont {Giardini}},
  \bibinfo {author} {\bibfnamefont {R.}~\bibnamefont {Giusteri}}, \bibinfo
  {author} {\bibfnamefont {C.}~\bibnamefont {Grimani}}, \bibinfo {author}
  {\bibfnamefont {J.}~\bibnamefont {Grzymisch}}, \bibinfo {author}
  {\bibfnamefont {I.}~\bibnamefont {Harrison}}, \bibinfo {author}
  {\bibfnamefont {G.}~\bibnamefont {Heinzel}}, \bibinfo {author} {\bibfnamefont
  {M.}~\bibnamefont {Hewitson}}, \bibinfo {author} {\bibfnamefont
  {D.}~\bibnamefont {Hollington}}, \bibinfo {author} {\bibfnamefont
  {D.}~\bibnamefont {Hoyland}}, \bibinfo {author} {\bibfnamefont
  {M.}~\bibnamefont {Hueller}}, \bibinfo {author} {\bibfnamefont
  {H.}~\bibnamefont {Inchausp\'e}}, \bibinfo {author} {\bibfnamefont
  {O.}~\bibnamefont {Jennrich}}, \bibinfo {author} {\bibfnamefont
  {P.}~\bibnamefont {Jetzer}}, \bibinfo {author} {\bibfnamefont
  {N.}~\bibnamefont {Karnesis}}, \bibinfo {author} {\bibfnamefont
  {B.}~\bibnamefont {Kaune}}, \bibinfo {author} {\bibfnamefont
  {N.}~\bibnamefont {Korsakova}}, \bibinfo {author} {\bibfnamefont {C.~J.}\
  \bibnamefont {Killow}}, \bibinfo {author} {\bibfnamefont {J.~A.}\
  \bibnamefont {Lobo}}, \bibinfo {author} {\bibfnamefont {I.}~\bibnamefont
  {Lloro}}, \bibinfo {author} {\bibfnamefont {L.}~\bibnamefont {Liu}}, \bibinfo
  {author} {\bibfnamefont {J.~P.}\ \bibnamefont {L\'opez-Zaragoza}}, \bibinfo
  {author} {\bibfnamefont {R.}~\bibnamefont {Maarschalkerweerd}}, \bibinfo
  {author} {\bibfnamefont {D.}~\bibnamefont {Mance}}, \bibinfo {author}
  {\bibfnamefont {N.}~\bibnamefont {Meshksar}}, \bibinfo {author}
  {\bibfnamefont {V.}~\bibnamefont {Mart\'{\i}n}}, \bibinfo {author}
  {\bibfnamefont {L.}~\bibnamefont {Martin-Polo}}, \bibinfo {author}
  {\bibfnamefont {J.}~\bibnamefont {Martino}}, \bibinfo {author} {\bibfnamefont
  {F.}~\bibnamefont {Martin-Porqueras}}, \bibinfo {author} {\bibfnamefont
  {I.}~\bibnamefont {Mateos}}, \bibinfo {author} {\bibfnamefont {P.~W.}\
  \bibnamefont {McNamara}}, \bibinfo {author} {\bibfnamefont {J.}~\bibnamefont
  {Mendes}}, \bibinfo {author} {\bibfnamefont {L.}~\bibnamefont {Mendes}},
  \bibinfo {author} {\bibfnamefont {M.}~\bibnamefont {Nofrarias}}, \bibinfo
  {author} {\bibfnamefont {S.}~\bibnamefont {Paczkowski}}, \bibinfo {author}
  {\bibfnamefont {M.}~\bibnamefont {Perreur-Lloyd}}, \bibinfo {author}
  {\bibfnamefont {A.}~\bibnamefont {Petiteau}}, \bibinfo {author}
  {\bibfnamefont {P.}~\bibnamefont {Pivato}}, \bibinfo {author} {\bibfnamefont
  {E.}~\bibnamefont {Plagnol}}, \bibinfo {author} {\bibfnamefont
  {J.}~\bibnamefont {Ramos-Castro}}, \bibinfo {author} {\bibfnamefont
  {J.}~\bibnamefont {Reiche}}, \bibinfo {author} {\bibfnamefont {D.~I.}\
  \bibnamefont {Robertson}}, \bibinfo {author} {\bibfnamefont {F.}~\bibnamefont
  {Rivas}}, \bibinfo {author} {\bibfnamefont {G.}~\bibnamefont {Russano}},
  \bibinfo {author} {\bibfnamefont {J.}~\bibnamefont {Slutsky}}, \bibinfo
  {author} {\bibfnamefont {C.~F.}\ \bibnamefont {Sopuerta}}, \bibinfo {author}
  {\bibfnamefont {T.}~\bibnamefont {Sumner}}, \bibinfo {author} {\bibfnamefont
  {D.}~\bibnamefont {Texier}}, \bibinfo {author} {\bibfnamefont {J.~I.}\
  \bibnamefont {Thorpe}}, \bibinfo {author} {\bibfnamefont {D.}~\bibnamefont
  {Vetrugno}}, \bibinfo {author} {\bibfnamefont {S.}~\bibnamefont {Vitale}},
  \bibinfo {author} {\bibfnamefont {G.}~\bibnamefont {Wanner}}, \bibinfo
  {author} {\bibfnamefont {H.}~\bibnamefont {Ward}}, \bibinfo {author}
  {\bibfnamefont {P.~J.}\ \bibnamefont {Wass}}, \bibinfo {author}
  {\bibfnamefont {W.~J.}\ \bibnamefont {Weber}}, \bibinfo {author}
  {\bibfnamefont {L.}~\bibnamefont {Wissel}}, \bibinfo {author} {\bibfnamefont
  {A.}~\bibnamefont {Wittchen}}, \ and\ \bibinfo {author} {\bibfnamefont
  {P.}~\bibnamefont {Zweifel}} (\bibinfo {collaboration} {LISA Pathfinder
  Collaboration}),\ }\bibfield  {title} {\enquote {\bibinfo {title} {Lisa
  pathfinder platform stability and drag-free performance},}\ }\href {\doibase
  10.1103/PhysRevD.99.082001} {\bibfield  {journal} {\bibinfo  {journal} {Phys.
  Rev. D}\ }\textbf {\bibinfo {volume} {99}},\ \bibinfo {pages} {082001}
  (\bibinfo {year} {2019})}\BibitemShut {NoStop}%
\bibitem [{\citenamefont {Koch}\ \emph {et~al.}(2018)\citenamefont {Koch},
  \citenamefont {Sanjuan}, \citenamefont {Gohlke}, \citenamefont {Mahrdt},
  \citenamefont {Brause}, \citenamefont {Braxmaier},\ and\ \citenamefont
  {Heinzel}}]{Koch2018}%
  \BibitemOpen
  \bibfield  {author} {\bibinfo {author} {\bibfnamefont {Alexander}\
  \bibnamefont {Koch}}, \bibinfo {author} {\bibfnamefont {Josep}\ \bibnamefont
  {Sanjuan}}, \bibinfo {author} {\bibfnamefont {Martin}\ \bibnamefont
  {Gohlke}}, \bibinfo {author} {\bibfnamefont {Christoph}\ \bibnamefont
  {Mahrdt}}, \bibinfo {author} {\bibfnamefont {Nils}\ \bibnamefont {Brause}},
  \bibinfo {author} {\bibfnamefont {Claus}\ \bibnamefont {Braxmaier}}, \ and\
  \bibinfo {author} {\bibfnamefont {Gerhard}\ \bibnamefont {Heinzel}},\
  }\bibfield  {title} {\enquote {\bibinfo {title} {Line of sight calibration
  for the laser ranging interferometer on-board the grace follow-on mission:
  on-ground experimental validation},}\ }\href {\doibase 10.1364/OE.26.025892}
  {\bibfield  {journal} {\bibinfo  {journal} {Optics Express}\ }\textbf
  {\bibinfo {volume} {26}},\ \bibinfo {pages} {25892} (\bibinfo {year}
  {2018})}\BibitemShut {NoStop}%
\bibitem [{\citenamefont {Morrison}\ \emph
  {et~al.}(1994{\natexlab{a}})\citenamefont {Morrison}, \citenamefont {Meers},
  \citenamefont {Robertson},\ and\ \citenamefont {Ward}}]{DWS1}%
  \BibitemOpen
  \bibfield  {author} {\bibinfo {author} {\bibfnamefont {E.}~\bibnamefont
  {Morrison}}, \bibinfo {author} {\bibfnamefont {B.~J.}\ \bibnamefont {Meers}},
  \bibinfo {author} {\bibfnamefont {D.~I.}\ \bibnamefont {Robertson}}, \ and\
  \bibinfo {author} {\bibfnamefont {H.}~\bibnamefont {Ward}},\ }\bibfield
  {title} {\enquote {\bibinfo {title} {Automatic alignment of optical
  interferometers},}\ }\href {\doibase 10.1364/AO.33.005041} {\bibfield
  {journal} {\bibinfo  {journal} {Applied Optics}\ }\textbf {\bibinfo {volume}
  {33}},\ \bibinfo {pages} {5041--5049} (\bibinfo {year}
  {1994}{\natexlab{a}})}\BibitemShut {NoStop}%
\bibitem [{\citenamefont {Morrison}\ \emph
  {et~al.}(1994{\natexlab{b}})\citenamefont {Morrison}, \citenamefont {Meers},
  \citenamefont {Robertson},\ and\ \citenamefont {Ward}}]{DWS2}%
  \BibitemOpen
  \bibfield  {author} {\bibinfo {author} {\bibfnamefont {E.}~\bibnamefont
  {Morrison}}, \bibinfo {author} {\bibfnamefont {B.~J.}\ \bibnamefont {Meers}},
  \bibinfo {author} {\bibfnamefont {D.~I.}\ \bibnamefont {Robertson}}, \ and\
  \bibinfo {author} {\bibfnamefont {H.}~\bibnamefont {Ward}},\ }\bibfield
  {title} {\enquote {\bibinfo {title} {Experimental demonstration of an
  automatic alignment system for optical interferometers},}\ }\href {\doibase
  10.1364/AO.33.005037} {\bibfield  {journal} {\bibinfo  {journal} {Applied
  Optics}\ }\textbf {\bibinfo {volume} {33}},\ \bibinfo {pages} {5037--5040}
  (\bibinfo {year} {1994}{\natexlab{b}})}\BibitemShut {NoStop}%
\bibitem [{\citenamefont {Gardner}(2005)}]{GARDNER}%
  \BibitemOpen
  \bibfield  {author} {\bibinfo {author} {\bibfnamefont {Floyd~M.}\
  \bibnamefont {Gardner}},\ }\href {\doibase 10.1002/0471732699} {\emph
  {\bibinfo {title} {Phaselock Techniques}}}\ (\bibinfo  {publisher} {John
  Wiley \& Sons},\ \bibinfo {address} {New York},\ \bibinfo {year}
  {2005})\BibitemShut {NoStop}%
\bibitem [{\citenamefont {Wanner}\ \emph {et~al.}(2012)\citenamefont {Wanner},
  \citenamefont {Heinzel}, \citenamefont {Kochkina}, \citenamefont {Mahrdt},
  \citenamefont {Sheard}, \citenamefont {Schuster},\ and\ \citenamefont
  {Danzmann}}]{DWSFAC}%
  \BibitemOpen
  \bibfield  {author} {\bibinfo {author} {\bibfnamefont {Gudrun}\ \bibnamefont
  {Wanner}}, \bibinfo {author} {\bibfnamefont {Gerhard}\ \bibnamefont
  {Heinzel}}, \bibinfo {author} {\bibfnamefont {Evgenia}\ \bibnamefont
  {Kochkina}}, \bibinfo {author} {\bibfnamefont {Christoph}\ \bibnamefont
  {Mahrdt}}, \bibinfo {author} {\bibfnamefont {Benjamin~S.}\ \bibnamefont
  {Sheard}}, \bibinfo {author} {\bibfnamefont {Soenke}\ \bibnamefont
  {Schuster}}, \ and\ \bibinfo {author} {\bibfnamefont {Karsten}\ \bibnamefont
  {Danzmann}},\ }\bibfield  {title} {\enquote {\bibinfo {title} {Methods for
  simulating the readout of lengths and angles in laser interferometers with
  gaussian beams},}\ }\href {\doibase 10.1016/j.optcom.2012.07.123} {\bibfield
  {journal} {\bibinfo  {journal} {Optics Communications}\ }\textbf {\bibinfo
  {volume} {285}},\ \bibinfo {pages} {4831--4839} (\bibinfo {year}
  {2012})}\BibitemShut {NoStop}%
\bibitem [{\citenamefont {Wanner}\ \emph {et~al.}(2015)\citenamefont {Wanner},
  \citenamefont {Schuster}, \citenamefont {Troebs},\ and\ \citenamefont
  {Heinzel}}]{SIGNALS}%
  \BibitemOpen
  \bibfield  {author} {\bibinfo {author} {\bibfnamefont {Gudrun}\ \bibnamefont
  {Wanner}}, \bibinfo {author} {\bibfnamefont {Soenke}\ \bibnamefont
  {Schuster}}, \bibinfo {author} {\bibfnamefont {Michael}\ \bibnamefont
  {Troebs}}, \ and\ \bibinfo {author} {\bibfnamefont {Gerhard}\ \bibnamefont
  {Heinzel}},\ }\bibfield  {title} {\enquote {\bibinfo {title} {A brief
  comparison of optical pathlength difference and various definitions for the
  interferometric phase},}\ }in\ \href {\doibase
  10.1088/1742-6596/610/1/012043} {\emph {\bibinfo {booktitle} {10th
  International LISA Symposium}}},\ \bibinfo {series} {Journal of Physics
  Conference Series}, Vol.\ \bibinfo {volume} {610}\ (\bibinfo {year}
  {2015})\BibitemShut {NoStop}%
\bibitem [{\citenamefont {Brause}(2018)}]{NILS}%
  \BibitemOpen
  \bibfield  {author} {\bibinfo {author} {\bibfnamefont {Nils~Christopher}\
  \bibnamefont {Brause}},\ }\emph {\bibinfo {title} {Auxiliary function
  development for the LISA metrology system}},\ \href
  {https://www.repo.uni-hannover.de/handle/123456789/3541} {Ph.D. thesis},\
  \bibinfo  {school} {Leibniz Universit\"at Hannover} (\bibinfo {year}
  {2018})\BibitemShut {NoStop}%
\bibitem [{\citenamefont {Chwalla}\ \emph {et~al.}(2020)\citenamefont
  {Chwalla}, \citenamefont {Danzmann}, \citenamefont {{\'{A}}lvarez},
  \citenamefont {Delgado}, \citenamefont {Barranco}, \citenamefont
  {Fitzsimons}, \citenamefont {Gerberding}, \citenamefont {Heinzel},
  \citenamefont {Killow}, \citenamefont {Lieser}, \citenamefont
  {Perreur-Lloyd}, \citenamefont {Robertson}, \citenamefont {Rohr},
  \citenamefont {Schuster}, \citenamefont {Schwarze}, \citenamefont
  {Tr\"{o}bs}, \citenamefont {Wanner},\ and\ \citenamefont
  {Ward}}]{Chwalla2020}%
  \BibitemOpen
  \bibfield  {author} {\bibinfo {author} {\bibfnamefont {M.}~\bibnamefont
  {Chwalla}}, \bibinfo {author} {\bibfnamefont {K.}~\bibnamefont {Danzmann}},
  \bibinfo {author} {\bibfnamefont {M.~Dovale}\ \bibnamefont {{\'{A}}lvarez}},
  \bibinfo {author} {\bibfnamefont {J.J.~Esteban}\ \bibnamefont {Delgado}},
  \bibinfo {author} {\bibfnamefont {G.~Fern{\'{a}}ndez}\ \bibnamefont
  {Barranco}}, \bibinfo {author} {\bibfnamefont {E.}~\bibnamefont
  {Fitzsimons}}, \bibinfo {author} {\bibfnamefont {O.}~\bibnamefont
  {Gerberding}}, \bibinfo {author} {\bibfnamefont {G.}~\bibnamefont {Heinzel}},
  \bibinfo {author} {\bibfnamefont {C.J.}\ \bibnamefont {Killow}}, \bibinfo
  {author} {\bibfnamefont {M.}~\bibnamefont {Lieser}}, \bibinfo {author}
  {\bibfnamefont {M.}~\bibnamefont {Perreur-Lloyd}}, \bibinfo {author}
  {\bibfnamefont {D.I.}\ \bibnamefont {Robertson}}, \bibinfo {author}
  {\bibfnamefont {J.M.}\ \bibnamefont {Rohr}}, \bibinfo {author} {\bibfnamefont
  {S.}~\bibnamefont {Schuster}}, \bibinfo {author} {\bibfnamefont {T.S.}\
  \bibnamefont {Schwarze}}, \bibinfo {author} {\bibfnamefont {M.}~\bibnamefont
  {Tr\"{o}bs}}, \bibinfo {author} {\bibfnamefont {G.}~\bibnamefont {Wanner}}, \
  and\ \bibinfo {author} {\bibfnamefont {H.}~\bibnamefont {Ward}},\ }\bibfield
  {title} {\enquote {\bibinfo {title} {Optical suppression of tilt-to-length
  coupling in the {LISA} long-arm interferometer},}\ }\href {\doibase
  10.1103/physrevapplied.14.014030} {\bibfield  {journal} {\bibinfo  {journal}
  {Physical Review Applied}\ }\textbf {\bibinfo {volume} {14}} (\bibinfo {year}
  {2020}),\ 10.1103/physrevapplied.14.014030}\BibitemShut {NoStop}%
\bibitem [{\citenamefont {Tr{\"o}bs}\ \emph {et~al.}(2018)\citenamefont
  {Tr{\"o}bs}, \citenamefont {Schuster}, \citenamefont {Lieser}, \citenamefont
  {Zwetz}, \citenamefont {Chwalla}, \citenamefont {Danzmann}, \citenamefont
  {Barr{\'a}nco}, \citenamefont {Fitzsimons}, \citenamefont {Gerberding},
  \citenamefont {Heinzel}, \citenamefont {Killow}, \citenamefont
  {Perreur-Lloyd}, \citenamefont {Robertson}, \citenamefont {Schwarze},
  \citenamefont {Wanner},\ and\ \citenamefont {Ward}}]{Trobs2018}%
  \BibitemOpen
  \bibfield  {author} {\bibinfo {author} {\bibfnamefont {M}~\bibnamefont
  {Tr{\"o}bs}}, \bibinfo {author} {\bibfnamefont {S}~\bibnamefont {Schuster}},
  \bibinfo {author} {\bibfnamefont {M}~\bibnamefont {Lieser}}, \bibinfo
  {author} {\bibfnamefont {M}~\bibnamefont {Zwetz}}, \bibinfo {author}
  {\bibfnamefont {M}~\bibnamefont {Chwalla}}, \bibinfo {author} {\bibfnamefont
  {K}~\bibnamefont {Danzmann}}, \bibinfo {author} {\bibfnamefont
  {G~Fern{\'a}ndez}\ \bibnamefont {Barr{\'a}nco}}, \bibinfo {author}
  {\bibfnamefont {E~D}\ \bibnamefont {Fitzsimons}}, \bibinfo {author}
  {\bibfnamefont {O}~\bibnamefont {Gerberding}}, \bibinfo {author}
  {\bibfnamefont {G}~\bibnamefont {Heinzel}}, \bibinfo {author} {\bibfnamefont
  {C~J}\ \bibnamefont {Killow}}, \bibinfo {author} {\bibfnamefont
  {M}~\bibnamefont {Perreur-Lloyd}}, \bibinfo {author} {\bibfnamefont {D~I}\
  \bibnamefont {Robertson}}, \bibinfo {author} {\bibfnamefont {T~S}\
  \bibnamefont {Schwarze}}, \bibinfo {author} {\bibfnamefont {G}~\bibnamefont
  {Wanner}}, \ and\ \bibinfo {author} {\bibfnamefont {H}~\bibnamefont {Ward}},\
  }\bibfield  {title} {\enquote {\bibinfo {title} {{Reducing tilt-to-length
  coupling for the LISA test mass interferometer}},}\ }\href
  {https://doi.org/10.1088/1361-6382/aab86c} {\bibfield  {journal} {\bibinfo
  {journal} {Classical and Quantum Gravity}\ }\textbf {\bibinfo {volume}
  {35}},\ \bibinfo {pages} {105001--22} (\bibinfo {year} {2018})}\BibitemShut
  {NoStop}%
\bibitem [{\citenamefont {{IfoCAD}}()}]{Ifocad}%
  \BibitemOpen
  \bibfield  {author} {\bibinfo {author} {\bibnamefont {{IfoCAD}}},\ }\href
  {http://www.lisa.aei-hannover.de/ifocad/} {\enquote {\bibinfo {title}
  {http://www.lisa.aei-hannover.de/ifocad/},}\ }\BibitemShut {NoStop}%
\bibitem [{\citenamefont {Wuchenich}\ \emph {et~al.}(2014)\citenamefont
  {Wuchenich}, \citenamefont {Mahrdt}, \citenamefont {Sheard}, \citenamefont
  {Francis}, \citenamefont {Spero}, \citenamefont {Miller}, \citenamefont
  {Mow-Lowry}, \citenamefont {Ward}, \citenamefont {Klipstein}, \citenamefont
  {Heinzel}, \citenamefont {Danzmann}, \citenamefont {McClelland},\ and\
  \citenamefont {Shaddock}}]{Wuchenich2014}%
  \BibitemOpen
  \bibfield  {author} {\bibinfo {author} {\bibfnamefont {Danielle M.~R.}\
  \bibnamefont {Wuchenich}}, \bibinfo {author} {\bibfnamefont {Christoph}\
  \bibnamefont {Mahrdt}}, \bibinfo {author} {\bibfnamefont {Benjamin~S.}\
  \bibnamefont {Sheard}}, \bibinfo {author} {\bibfnamefont {Samuel~P.}\
  \bibnamefont {Francis}}, \bibinfo {author} {\bibfnamefont {Robert~E.}\
  \bibnamefont {Spero}}, \bibinfo {author} {\bibfnamefont {John}\ \bibnamefont
  {Miller}}, \bibinfo {author} {\bibfnamefont {Conor~M.}\ \bibnamefont
  {Mow-Lowry}}, \bibinfo {author} {\bibfnamefont {Robert~L.}\ \bibnamefont
  {Ward}}, \bibinfo {author} {\bibfnamefont {William~M.}\ \bibnamefont
  {Klipstein}}, \bibinfo {author} {\bibfnamefont {Gerhard}\ \bibnamefont
  {Heinzel}}, \bibinfo {author} {\bibfnamefont {Karsten}\ \bibnamefont
  {Danzmann}}, \bibinfo {author} {\bibfnamefont {David~E.}\ \bibnamefont
  {McClelland}}, \ and\ \bibinfo {author} {\bibfnamefont {Daniel~A.}\
  \bibnamefont {Shaddock}},\ }\bibfield  {title} {\enquote {\bibinfo {title}
  {Laser link acquisition demonstration for the grace follow-on mission},}\
  }\href {\doibase 10.1364/OE.22.011351} {\bibfield  {journal} {\bibinfo
  {journal} {Opt. Express}\ }\textbf {\bibinfo {volume} {22}},\ \bibinfo
  {pages} {11351--11366} (\bibinfo {year} {2014})}\BibitemShut {NoStop}%
\bibitem [{\citenamefont {Koch}(2020)}]{Koch2020}%
  \BibitemOpen
  \bibfield  {author} {\bibinfo {author} {\bibfnamefont {Alexander}\
  \bibnamefont {Koch}},\ }\emph {\bibinfo {title} {Link Acquisition and
  Optimization for Intersatellite Laser Interferometry}},\ \href
  {https://www.repo.uni-hannover.de/handle/123456789/9856} {Ph.D. thesis},\
  \bibinfo  {school} {Leibniz Universit\"at Hannover} (\bibinfo {year}
  {2020})\BibitemShut {NoStop}%
\bibitem [{\citenamefont {Mow-Lowry}\ and\ \citenamefont
  {Martynov}(2019)}]{MowLowry2019}%
  \BibitemOpen
  \bibfield  {author} {\bibinfo {author} {\bibfnamefont {C~M}\ \bibnamefont
  {Mow-Lowry}}\ and\ \bibinfo {author} {\bibfnamefont {D}~\bibnamefont
  {Martynov}},\ }\bibfield  {title} {\enquote {\bibinfo {title} {A 6d
  interferometric inertial isolation system},}\ }\href {\doibase
  10.1088/1361-6382/ab4e01} {\bibfield  {journal} {\bibinfo  {journal}
  {Classical and Quantum Gravity}\ }\textbf {\bibinfo {volume} {36}},\ \bibinfo
  {pages} {245006} (\bibinfo {year} {2019})}\BibitemShut {NoStop}%
\bibitem [{\citenamefont {Schlamminger}\ \emph {et~al.}(2008)\citenamefont
  {Schlamminger}, \citenamefont {Choi}, \citenamefont {Wagner}, \citenamefont
  {Gundlach},\ and\ \citenamefont {Adelberger}}]{Schlamminger2008}%
  \BibitemOpen
  \bibfield  {author} {\bibinfo {author} {\bibfnamefont {S.}~\bibnamefont
  {Schlamminger}}, \bibinfo {author} {\bibfnamefont {K.-Y.}\ \bibnamefont
  {Choi}}, \bibinfo {author} {\bibfnamefont {T.~A.}\ \bibnamefont {Wagner}},
  \bibinfo {author} {\bibfnamefont {J.~H.}\ \bibnamefont {Gundlach}}, \ and\
  \bibinfo {author} {\bibfnamefont {E.~G.}\ \bibnamefont {Adelberger}},\
  }\bibfield  {title} {\enquote {\bibinfo {title} {Test of the equivalence
  principle using a rotating torsion balance},}\ }\href {\doibase
  10.1103/physrevlett.100.041101} {\bibfield  {journal} {\bibinfo  {journal}
  {Physical Review Letters}\ }\textbf {\bibinfo {volume} {100}} (\bibinfo
  {year} {2008}),\ 10.1103/physrevlett.100.041101}\BibitemShut {NoStop}%
\bibitem [{\citenamefont {Luo}\ \emph {et~al.}(2009)\citenamefont {Luo},
  \citenamefont {Liu}, \citenamefont {Tu}, \citenamefont {Shao}, \citenamefont
  {Liu}, \citenamefont {Yang}, \citenamefont {Li},\ and\ \citenamefont
  {Zhang}}]{Luo2009}%
  \BibitemOpen
  \bibfield  {author} {\bibinfo {author} {\bibfnamefont {Jun}\ \bibnamefont
  {Luo}}, \bibinfo {author} {\bibfnamefont {Qi}~\bibnamefont {Liu}}, \bibinfo
  {author} {\bibfnamefont {Liang-Cheng}\ \bibnamefont {Tu}}, \bibinfo {author}
  {\bibfnamefont {Cheng-Gang}\ \bibnamefont {Shao}}, \bibinfo {author}
  {\bibfnamefont {Lin-Xia}\ \bibnamefont {Liu}}, \bibinfo {author}
  {\bibfnamefont {Shan-Qing}\ \bibnamefont {Yang}}, \bibinfo {author}
  {\bibfnamefont {Qing}\ \bibnamefont {Li}}, \ and\ \bibinfo {author}
  {\bibfnamefont {Ya-Ting}\ \bibnamefont {Zhang}},\ }\bibfield  {title}
  {\enquote {\bibinfo {title} {Determination of the newtonian gravitational
  {ConstantGwith} time-of-swing method},}\ }\href {\doibase
  10.1103/physrevlett.102.240801} {\bibfield  {journal} {\bibinfo  {journal}
  {Physical Review Letters}\ }\textbf {\bibinfo {volume} {102}} (\bibinfo
  {year} {2009}),\ 10.1103/physrevlett.102.240801}\BibitemShut {NoStop}%
\bibitem [{\citenamefont {Li}\ \emph {et~al.}(2018)\citenamefont {Li},
  \citenamefont {Xue}, \citenamefont {Liu}, \citenamefont {Wu}, \citenamefont
  {Yang}, \citenamefont {Shao}, \citenamefont {Quan}, \citenamefont {Tan},
  \citenamefont {Tu}, \citenamefont {Liu}, \citenamefont {Xu}, \citenamefont
  {Liu}, \citenamefont {Wang}, \citenamefont {Hu}, \citenamefont {Zhou},
  \citenamefont {Luo}, \citenamefont {Wu}, \citenamefont {Milyukov},\ and\
  \citenamefont {Luo}}]{Li2018}%
  \BibitemOpen
  \bibfield  {author} {\bibinfo {author} {\bibfnamefont {Qing}\ \bibnamefont
  {Li}}, \bibinfo {author} {\bibfnamefont {Chao}\ \bibnamefont {Xue}}, \bibinfo
  {author} {\bibfnamefont {Jian-Ping}\ \bibnamefont {Liu}}, \bibinfo {author}
  {\bibfnamefont {Jun-Fei}\ \bibnamefont {Wu}}, \bibinfo {author}
  {\bibfnamefont {Shan-Qing}\ \bibnamefont {Yang}}, \bibinfo {author}
  {\bibfnamefont {Cheng-Gang}\ \bibnamefont {Shao}}, \bibinfo {author}
  {\bibfnamefont {Li-Di}\ \bibnamefont {Quan}}, \bibinfo {author}
  {\bibfnamefont {Wen-Hai}\ \bibnamefont {Tan}}, \bibinfo {author}
  {\bibfnamefont {Liang-Cheng}\ \bibnamefont {Tu}}, \bibinfo {author}
  {\bibfnamefont {Qi}~\bibnamefont {Liu}}, \bibinfo {author} {\bibfnamefont
  {Hao}\ \bibnamefont {Xu}}, \bibinfo {author} {\bibfnamefont {Lin-Xia}\
  \bibnamefont {Liu}}, \bibinfo {author} {\bibfnamefont {Qing-Lan}\
  \bibnamefont {Wang}}, \bibinfo {author} {\bibfnamefont {Zhong-Kun}\
  \bibnamefont {Hu}}, \bibinfo {author} {\bibfnamefont {Ze-Bing}\ \bibnamefont
  {Zhou}}, \bibinfo {author} {\bibfnamefont {Peng-Shun}\ \bibnamefont {Luo}},
  \bibinfo {author} {\bibfnamefont {Shu-Chao}\ \bibnamefont {Wu}}, \bibinfo
  {author} {\bibfnamefont {Vadim}\ \bibnamefont {Milyukov}}, \ and\ \bibinfo
  {author} {\bibfnamefont {Jun}\ \bibnamefont {Luo}},\ }\bibfield  {title}
  {\enquote {\bibinfo {title} {Measurements of the gravitational constant using
  two independent methods},}\ }\href {\doibase 10.1038/s41586-018-0431-5}
  {\bibfield  {journal} {\bibinfo  {journal} {Nature}\ }\textbf {\bibinfo
  {volume} {560}},\ \bibinfo {pages} {582--588} (\bibinfo {year}
  {2018})}\BibitemShut {NoStop}%
\end{thebibliography}

%

\end{document}